\documentclass[pdflatex,sn-mathphys-ay]{sn-jnl}
\usepackage{graphicx}
\usepackage{amsmath}
\usepackage{amssymb}

\DeclareMathOperator*{\argmax}{arg\,max}
\newtheorem{lemma}{Lemma}

\begin{document}

\title{Post-selection inference for quantifying uncertainty in changes in variance}

\author*{\fnm{Rachel} \sur{Carrington*}}\email{rjc204@bath.ac.uk}
\author{\fnm{Paul} \sur{Fearnhead}}\email{p.fearnhead@lanfcaster.ac.uk}

\affil{\orgdiv{School of Mathematical Sciences}, \orgname{Lancaster University}, \country{UK}}

\abstract{Quantifying uncertainty in detected changepoints is an important problem. However it is challenging as the naive approach would use the data twice, first to detect the changes, and then to test them. This will bias the test, and can lead to anti-conservative $p$-values. One approach to avoid this is to use ideas from post-selection inference, which conditions on the information in the data used to choose which changes to test. As a result this produces valid $p$-values; that is, $p$-values that have a uniform distribution if there is no change. Currently such methods have been developed for detecting changes in mean only. This paper presents two approaches for constructing post-selection $p$-values for detecting changes in variance. These vary depending on the method use to detect the changes, but are general in terms of being applicable for a range of change-detection methods and a range of hypotheses that we may wish to test.}


\maketitle

\section{Introduction} \label{sec:intro}

The problem of detecting changes in time series has received a great deal of attention in recent years, with applications in finance \citep{schroder2013adaptive},  quality control \citep{amiri2015probabilistic}, climate modelling \citep{reeves2007review, shi2022changepoint}, genome sequencing \citep{muggeo2011efficient, caron2012line}, neuroscience \citep{anastasiou2022new}, and epidemic modelling \citep{jiang2023time}, amongst many others. 
Various methods have been developed for detecting different types of change, for example changes in mean, variance, and slope.
These are often developed based on developing a test for a single change of a specific type. To detect multiple changes, this test can be used recursively within algorithms such as binary segmentation \citep{scott1974scott} and its variants such as wild binary segmentation \citep{fryzlewicz2014wild}, seeded binary segmentation \citep{kovacs2023seeded}, and narrowest over threshold \citep{baranowski2019narrowest}. Alternatively one can write down a likelihood for the data as a function of the number and location of changes and aim to estimate the changes by minimising a penalised version of this likelihood -- for many models this optimisation can be performed exactly, see \cite{killick2012optimal, maidstone2017optimal}.
See \cite{fearnhead2020relating} and \cite{shi2022comparison} for recent reviews of changepoint methods.

Recently, increasing attention has been given to the problem of estimating the uncertainty associated with detected changepoints \cite[]{frick2014multiscale,li2016fdr,chen2023quantifying}. This is challenging due to the interplay of different elements of uncertainty, in terms of the number and location of the changepoints. Bayesian approaches \cite[]{barry1993bayesian,fearnhead2006exact,nam2012quantifying,cappello2023bayesian} give a natural way of quantifying this uncertainty, but require specifying prior information with the results sensitive to these choices. Alternatively one can use the properties of test statistics under the assumption of no changepoints. For example, \cite{fryzlewicz2023narrowest} and \cite{fryzlewicz2024robust} use a probabilistic bound on the maximum of the test statistic for the presence of a change over all possible intervals of data to find  the narrowest set of intervals which each must contain at least one changepoint with a certain probability. Similar bounds are used within the MOSUM method \citep{meier2021mosum} to produce asymptotic $p$-values for detected changepoints \citep{eichinger2018mosum}.

In this paper we consider post-selection inference methods \cite[]{bayarri1999quantifying, kuchibhotla2022post} for quantifying the degree of uncertainty associated with detected changepoints. The idea is to construct a $p$-value for each detected changepoint, with these $p$-values being valid in finite samples, albeit under strong distributional assumptions for the data. This is a challenging problem in general as, if we were to use the data to estimate changepoint locations, naively re-using the same data to estimate the uncertainty associated with each changepoint will lead to biased results. The idea of post-selection inference is, when testing for the presence of a changepoint at some detected location $\tau$, we condition on the information in the data that led to detection at $\tau$ when calculating the $p$-value. To date, almost all work in this area has focused on the univariate change in mean case, e.g. \citep{hyun2021post, jewell2022testing, duy2022more, carrington2023improving}, or the piecewise constant polynomial model \citep{mehrizi2021valid}. 
As far as we know, there are no current approaches for quantifying the evidence for detected changes in variance. In this paper we fill this gap by extending the post-selection inference ideas to cover the change in variance setting.

The post-selection inference methods for the change in mean problem use the fact that the natural test statistic for a change at a given location will have a normal distribution under the null. To account for the fact that we are using the same data to calculate the test statistic as to detect changes -- and thus to decide which locations to test -- we need to condition this distribution on the fact that the tested location is a detected changepoint. This leads to a conditional null distribution that is constrained to lie on an interval \citep{hyun2021post} or a union of intervals \citep{jewell2022testing} of the real line -- depending on precisely what information is conditioned on. Importantly it is straightforward to calculate this interval or intervals for methods for detecting a change in mean that use the popular CUSUM statistic within a binary segmentation algorithm or a penalised likelihood approach. One of the advantages of this approach is that it is general: it applies to a wide range of methods for detecting changes, and also to most natural null hypotheses that one may wish to test.

In this paper, we consider the two most common ways of detecting a change in variance. The first uses the idea that a change in variance in some time series, $X_t$ say, will correspond to a change in mean in $X_t^2$. Thus we can shift the data so it has mean zero, and apply a method for detecting a change in mean to the square of this shifted data. The second method is where we estimate changepoint locations by directly maximizing the likelihood of the piecewise constant variance model. In both cases we then use the same test statistic for testing whether a detected change is real. This test statistic is equivalent to the likelihood-ratio test statistic for a change in variance in Gaussian data, and is the proportion of the sum of squares of the data in the region prior to the putative change. If we had used separate information, or independent data, to choose which putative change location to test then the null distribution of this test statistic is a Beta distribution with known parameters. For our setting we need to calculate the null distribution conditional on detecting the change, which leads to constraining the Beta distribution to a union of intervals of $[0,1]$. 

To calculate the post-selection $p$-values requires us to calculate these intervals. For the first approach to detecting changes, this can be done analytically using the same ideas as for the change in mean setting. For the second approach, this is no longer possible so we show how Monte Carlo can be used to approximate the $p$-values \cite[see][for similar use of Monte Carlo in post-selection inference]{saha2022inferring}.

The rest of the paper is organized as follows. In Section \ref{sec:test} we define the model and set up the test we will use to detect changes in variance. Section \ref{sec:methods} then describes the methods we will use to calculate $p$-values. Section \ref{sec:ext} details some extensions to different scenarios, such as different null hypotheses and changepoint detection methods. Section \ref{sec:sims} presents results of our methods on simulated data, and in Section \ref{sec:real} we implement our method on a financial data set. Section \ref{sec:discussion} then concludes with some discussion of our results and suggestions for further research.

\section{Testing for a change in variance} \label{sec:test}

We consider the univariate Gaussian model with fixed mean and piecewise constant variance:
\begin{equation}
    X_t = \mu + \sigma_t \epsilon_t, ~~ t = 1, \ldots, T,
  \label{eq:var-model}
\end{equation}
where $\mu$ is assumed known, $\epsilon_t \sim N(0,1)$, and $\sigma_{t + 1} = \sigma_t$ except at $K$ changepoints $\{\tau_1, \ldots, \tau_K\}$.

Given the model in \eqref{eq:var-model}, suppose that we have a data set $\boldsymbol{X} = \left( X_1, \ldots, X_T \right)$ and we have obtained a set of estimated changepoint locations $\mathcal{M}(\boldsymbol{X}) = \{\hat{\tau}_1, \ldots, \hat{\tau}_K\}$, such that $0 = \hat{\tau}_0 < \hat{\tau}_1 < \ldots < \hat{\tau}_K < \hat{\tau}_{K+1} = T$. For a particular changepoint $\hat{\tau}$, we want to test whether there is indeed a changepoint at, or close to, $\hat{\tau}$.

First, we need to define our null hypothesis and choose a test statistic. Since we are generally interested in whether there is a change close to $\hat{\tau}$, rather than exactly at $\hat{\tau}$, it makes sense to set the null hypothesis to be that there is no changepoint within a certain window of $\hat{\tau}$. There are various ways of choosing such a window, which will generally be informed by our particular application: for example, we can use a fixed window length, or allow the window to be determined by the locations of neighbouring changepoint estimates. For now we will assume that we have some fixed $h$ such that the window is $(\hat{\tau} - h, \hat{\tau} + h)$, but note that the methods we describe are easily extended to other choices of $H_0$; we will discuss this in more detail in Section \ref{sec:ext-1}.

Hence, we have
\begin{equation}
  \label{eq:h0}
    H_0 ~ : ~ \sigma_{\hat{\tau} - h + 1}^2 = \ldots = \sigma_{\hat{\tau}}^2 = \sigma_{\hat{\tau} + 1}^2 = \ldots = \sigma_{\hat{\tau} + h}^2,
\end{equation}
with $H_1$ being that there is at least one inequality. 

Let $C_{t_1:t_2}^2 = \sum_{t = t_1}^{t_2} X_t^2$.
A natural choice of test statistic is to consider what proportion of the total sum of squares within the region $(\hat{\tau} - h, \hat{\tau} + h)$ is accounted for by the data before the changepoint, i.e.
\begin{equation*}
    \phi = \frac{C_{(\hat{\tau} - h + 1):\hat{\tau}}^2}{C_{(\hat{\tau} - h + 1):(\hat{\tau} + h)}^2} = \frac{C_{(\hat{\tau} - h + 1):\hat{\tau}}^2}{C_{(\hat{\tau} - h + 1):\hat{\tau}}^2 + C_{(\hat{\tau} + 1):(\hat{\tau} + h)}^2}.
\end{equation*}
The terms in the last expression, $C_{(\hat{\tau} - h + 1):\hat{\tau}}^2$ and $C_{(\hat{\tau} + 1):(\hat{\tau} + h)}^2$, are independent gamma random variables with shape parameter $h/2$ and scale parameter $2\sigma^2$ under $H_0$. Thus,  under $H_0$, $\phi \sim \text{Beta} \left( \frac{h}{2}, \frac{h}{2} \right)$. Under $H_1$, we would expect $\phi$ to be closer to $0$ if the variance increases after $\hat{\tau}$, and closer to $1$ if $\hat{\tau}$ is followed by a decrease in variance.

Our test statistic, $\phi$, is equivalent to the F statistic, $F = \frac{C^2_{(\hat{\tau} - h + 1):\hat{\tau}}}{C^2_{(\hat{\tau} + 1):(\hat{\tau} + h)}}$, a standard test statistic used to test for a change in variance, in the sense that each of $\phi$ and $F$ can be derived from the other ($\phi = \frac{F}{1 + F}$ and $F = \frac{\phi}{1 - \phi}$). Hence using either will lead to the same result. The formulation we define above is more convenient, as it allows us to write $X_t^2$ as a linear function of $\phi$ (see below), which will simplify the calculations in the next section.

Let $\phi_{obs}$ denote the observed value of $\phi$. If $\hat{\tau}$ corresponds to an increase in variance, the one-sided $p$-value is $p = \Pr_{H_0} (\phi \leq \phi_{obs})$; if $\hat{\tau}$ corresponds to a decrease in variance, then $p = \Pr_{H_0} (\phi \geq \phi_{obs})$.
To calculate the two-sided $p$-value, we need to first calculate $\phi_*$ such that
\begin{equation*}
    \Pr_{H_0} \left( \phi \leq \phi_* \right) = 1 - \Pr_{H_0} \left( \phi \leq \phi_{obs} \right).
\end{equation*}
Then, we let $\phi_{lower} = \min \{\phi_{obs}, \phi_*\}$ and $\phi_{upper} = \max \{\phi_{obs}, \phi_*\}$, and the two-sided $p$-value is
\begin{equation}
    \label{eq:pvalue}
    p = \Pr_{H_0} \left( \phi \leq \phi_{lower} \text{ or } \phi \geq \phi_{upper} \right).
\end{equation}
We will assume for the remainder of this paper that we are using the two-sided $p$-value, i.e., we are not concerned about the direction of the change. However, our methods apply equally to the one-sided $p$-value case.

\subsection{Post-selection inference} \label{sec:inf}

\begin{figure}
    \centering
    \includegraphics[width=0.5\linewidth]{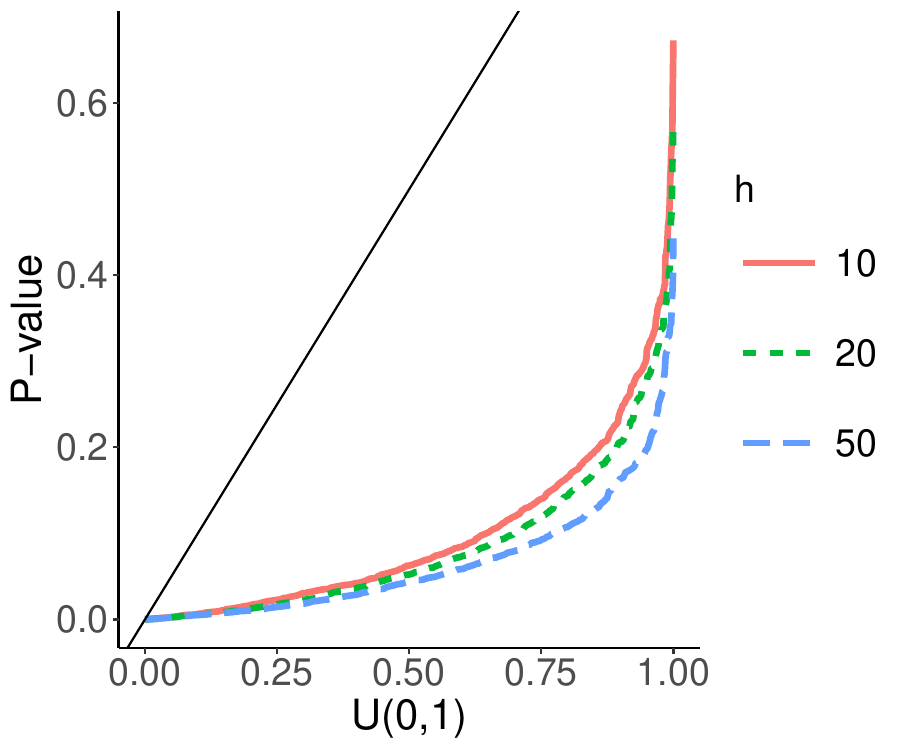}
    \caption{\small{QQ plot of $p$-values, calculated using \eqref{eq:pvalue}, for data simulated under the null hypothesis of no change, where we selected change locations to test from running binary segmentation on the same data. Different coloured lines correspond to different choices of $h$ in the null hypothesis. Re-using the data results in $p$-values that are much smaller than they should be.}}
    \label{fig:p-unconditional}
\end{figure}

The $p$-value in \eqref{eq:pvalue} is valid if we use separate information, or independent data, to calculate the $p$-value from that which we use to choose the location to test. However, if we use the same data to calculate the $p$-value as we have used to estimate changepoint locations, this introduces bias into the $p$-value. In particular, as demonstrated by Figure \ref{fig:p-unconditional}, we tend to obtain $p$-values that are much smaller than they should be, so the Type I error rate is too high; this is because in general we have chosen to test only at the locations where the test statistic is close to either 0 or 1, resulting in $p$-values that are too small. In order to get valid $p$-values in this case, we need to calculate the probability of \eqref{eq:pvalue} conditional on the information used to choose to test for a change at $\hat{\tau}$: here, this is the fact that $\hat{\tau}$ is in the set of estimated changepoints. Letting $\mathcal{M}(\boldsymbol{X})$ denote the set of estimated changepoints $\{\hat{\tau}_1, \ldots, \hat{\tau}_K\}$ which we obtain when we apply a particular changepoint algorithm to $\boldsymbol{X}$, we therefore want to calculate the $p$-value in \eqref{eq:pvalue} conditional on the fact that $\hat{\tau} \in \mathcal{M}(\boldsymbol{X})$:
\begin{equation*}
    \Pr_{H_0} \left( \phi \leq \phi_{lower} \text{ or } \phi \geq \phi_{upper} ~ | ~ \hat{\tau} \in \mathcal{M}(\boldsymbol{X}) \right).
\end{equation*}

Our null hypothesis \eqref{eq:h0} is that the variance is constant inside the region $\{\hat{\tau} - h + 1, \ldots, \hat{\tau} + h\}$. This does not specify the value that $\sigma_t^2$ takes for any $t$ outside this window, nor the value of the fixed variance within the window. Therefore, in order for the test to be well-defined, i.e. have the same distribution for all data generating processes consistent with the null hypothesis and the model, we must condition on sufficient statistics for these parameters: the values of $X_t$ outside of $\{\hat{\tau} - h + 1, \ldots, \hat{\tau} + h\}$, and the sum of squares within the window: $C_{(\hat{\tau} - h + 1):(\hat{\tau} + h)}^2 = \sum_{t=\hat{\tau} - h + 1}^{\hat{\tau} + h} X_t^2$.

To make calculating the $p$-values easier it is common to additionally condition on the data within the region we are testing that is orthogonal to $\phi$. We can condition on extra information like this whilst still retaining valid $p$-values (i.e. uniform $p$-values under $H_0$), but it may result in some loss of power \cite[]{fithian2014optimal,jewell2022testing}. In Section \ref{sec:sims-increasing-power}, we consider reducing this conditioning.

\begin{figure}
    \centering
    \includegraphics[width=0.8\linewidth]{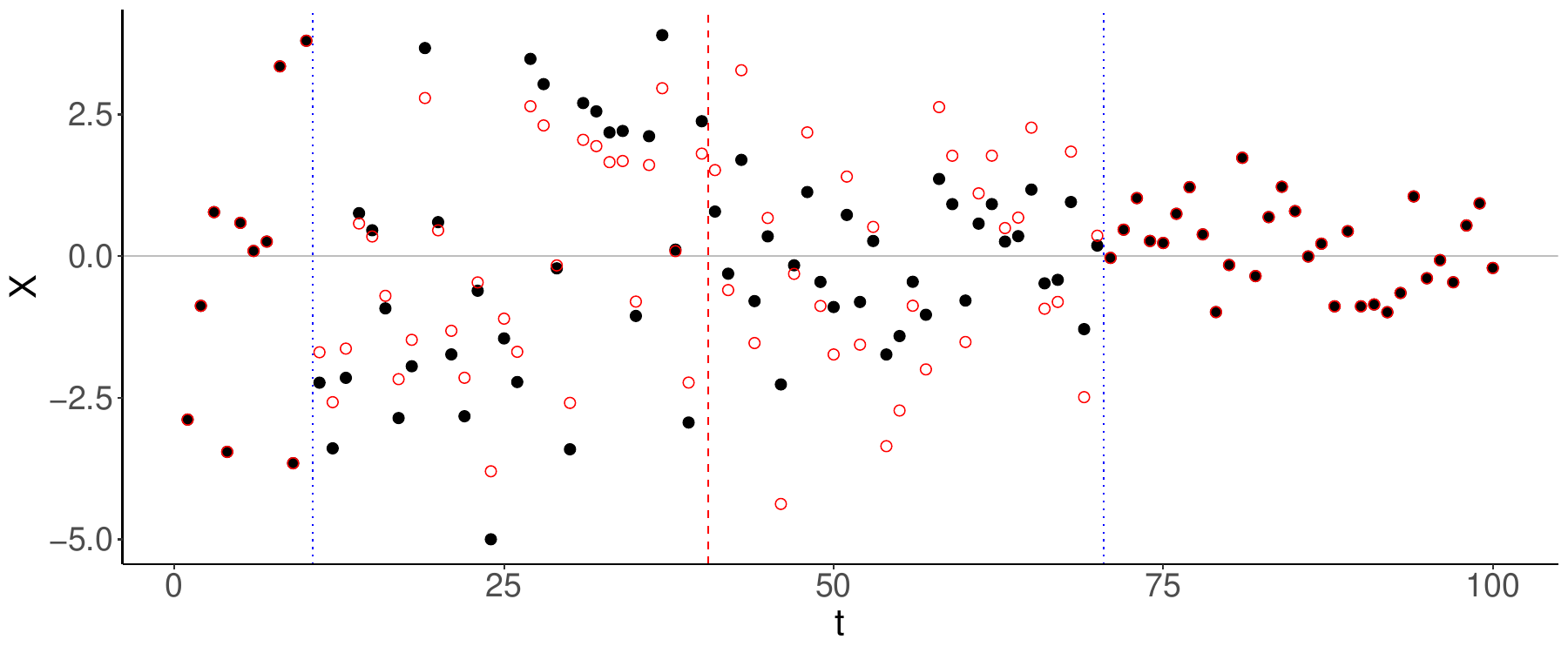}
    \caption{\small{An example of how the data is transformed when we apply the transformation $\boldsymbol{X}'(\phi)$. Here, the black points (filled circles) correspond to $\boldsymbol{X}$ and the red points (empty circles) to $\boldsymbol{X}'(0.5)$. Since $\phi_{obs} > 0.5$ in this case, the points before $\hat{\tau}$ (shown by the dashed red line) are pulled in towards 0, and the points after the changepoint move away from 0. The overall sum of squares in the region $\{\hat{\tau} - h + 1, \ldots, \hat{\tau} + h\}$ remains constant.}}
    \label{fig:x-phi}
\end{figure}

To make explicit the dependence of $\boldsymbol{X}$ on $\phi$, we consider a re-parameterization of $\{X_{\hat{\tau} - h + 1}, \ldots, X_{\hat{\tau} + h}\}$. Let $C_0^2 = C_{(\hat{\tau} - h + 1):(\hat{\tau} + h)}^2$, the sum of the square of the data in the region being tested, and let
\begin{equation*}
    \begin{split}
        W_i^l & = \frac{C_{(\hat{\tau} - h + 1):(\hat{\tau} - h + i)}^2}{C_{(\hat{\tau} - h + 1):(\hat{\tau} - h + i + 1)}^2}, ~~ i = 1, \ldots, h - 1; \\
        W_i^r & = \frac{C_{(\hat{\tau} + 1):(\hat{\tau} + i)}^2}{C_{(\hat{\tau} + 1):(\hat{\tau} + i + 1)}^2}, ~~ i = 1, \ldots, h - 1.
    \end{split}
\end{equation*}

The following results show how we can write the square of $\{X_{\hat{\tau} - h + 1}, \ldots, X_{\hat{\tau} + h}\}$ in terms of $\phi$, $C^2_0$, and $W^r_i$ and $W^l_i$ for $i = 1, \ldots, h - 1$, and that the latter set of variables are independent under the null hypothesis.
\begin{lemma}
  \label{lem:X}
    We can write each $X_t$, for $t = \hat{\tau} - h + 1, \ldots, \hat{\tau} + h$, in terms of $W_1^l, \ldots, W_{h-1}^l$, $W_1^r, \ldots, W_{h-1}^r$, $\phi$, and $C_0^2$. More specifically, for $i = 1, \ldots, h$, we can write
    \begin{equation}
      \label{eq:W}
        \begin{split}
            X_{\hat{\tau} - h + i}^2 & = \begin{cases}
                (1 - W_{i-1}^l) \prod_{k=i}^{h-1} W_k^l C_0^2 \phi & i = 1, \ldots, h - 1 \\
                (1 - W_{i-1}^l) C_0^2 \phi & i = h
            \end{cases}  \\
            X_{\hat{\tau} + i}^2 & = \begin{cases}
                (1 - W_{i-1}^r) \prod_{k=i}^{h-1} W_k^r C_0^2 (1 - \phi) & i = 1, \ldots, h - 1 \\
                (1 - W_{i-1}^r) C_0^2 (1 - \phi) & i = h,
            \end{cases}
        \end{split}
    \end{equation}
    where for convenience we set $W_0^l = W_0^r = 0$.
\end{lemma}

\begin{lemma}
  \label{lem:independence}
    Under $H_0$, $W_1^l, \ldots, W_{h-1}^l, W_1^r, \ldots, W_{h-1}^r, \phi, C_0^2$ are all independent.
\end{lemma}

Lemma \ref{lem:X} is easily verified. The proof of Lemma \ref{lem:independence} is given in the supplementary material.

Therefore, conditioning on the dimensions of the data which are orthogonal to $\phi$ is equivalent to conditioning on the values of $W_1^l, \ldots, W_{h-1}^l$, $W_1^r, \ldots, W_{h-1}^r$, and $C_0^2$. We can see from \eqref{eq:W} above that doing this while allowing only $\phi$ to change has the effect of multiplying $X_{\hat{\tau} - h + 1, \ldots, X_{\hat{\tau}}}$ by $\sqrt{\frac{\phi}{\phi_{obs}}}$ and multiplying $X_{\hat{\tau} + 1} , \ldots, X_{\hat{\tau} + h}$ by $\sqrt{\frac{1 - \phi}{1 - \phi_{obs}}}$. As we change $\phi$ we are re-scaling the data before and after $\hat{\tau}$ so that the proportion of the sum of squares of the data before and after $\hat{\tau}$ changes whilst the overall sum of squares of the data remains unchanged.

Hence, let
\begin{equation}
  \label{eq:x-prime}
    X_t'(\phi) = \begin{cases}
        X_t & t \leq \hat{\tau} - h \text{ or } t \geq \hat{\tau} + h + 1 \\
        \sqrt{\frac{\phi}{\phi_{obs}}} X_t & \hat{\tau} - h + 1 \leq t \leq \hat{\tau} \\
        \sqrt{\frac{1 - \phi}{1 - \phi_{obs}}} X_t & \hat{\tau} + 1 \leq t \leq \hat{\tau} + h.
    \end{cases}
\end{equation}
We consider only the subset of possible data sets $\{\boldsymbol{X}'(\phi) : \phi \in \mathbb{R}\}$. Letting $\mathcal{S}$ denote the set $\mathcal{S} = \{\phi : \hat{\tau} \in \mathcal{M}(\boldsymbol{X}'(\phi))\}$ -- the set of $\phi$ values such that the changepoint of interest, $\hat{\tau}$, is included as a change in the output of the changepoint algorithm -- the post-selection $p$-value we want to calculate is
\begin{equation}
    p = \Pr \left( \phi \leq \phi_{lower} \text{ or } \phi \geq \phi_{upper} ~ | ~ \phi \in \mathcal{S} \right).
  \label{eq:pval-con}
\end{equation}
Since we know that, under $H_0$, $\phi \sim \text{Beta} \left( \frac{h}{2}, \frac{h}{2} \right)$, the conditional distribution of $\phi | \phi \in \mathcal{S}$ is $\text{Beta} \left( \frac{h}{2}, \frac{h}{2} \right)$ truncated to $\mathcal{S}$. Hence, calculating $p$ depends on calculating $\mathcal{S}$. Calculation of $\mathcal{S}$ depends on the method used to estimate changepoints. In the following sections we describe how to do this for two classes of methods: in Section \ref{sec:cusum}, we cover approaches based on finding changes in mean in $X_t^2$ using cumulative sums of squares, whilst in Section \ref{sec:lrs} we develop a more general approach, focusing on changepoint detection methods which attempt to maximize the likelihood ratio statistic.

\section{Methods} \label{sec:methods}

\subsection{CUSUM for change in variance} \label{sec:cusum}

\begin{figure}
    \centering
    \includegraphics[width=0.98\linewidth]{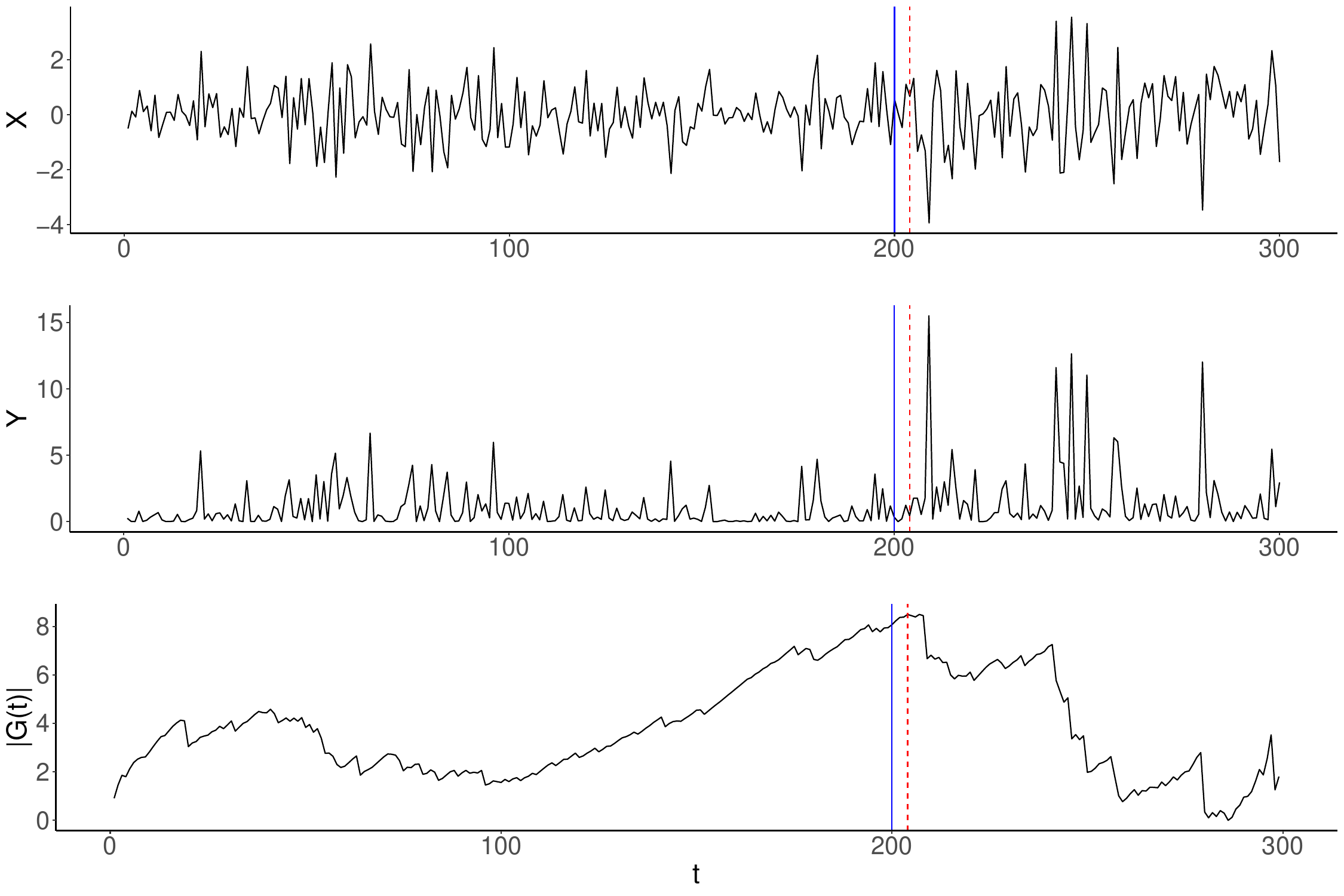}
    \caption{\small{A simulated data set with one change in variance. The data are simulated from $N(0, 1)$ for $t = 1, \ldots, 200$ and from $N(0, 1.7)$ for $t = 201, \ldots, 300$. The blue line shows the true changepoint location and the dashed red line the changepoint location estimated by maximizing $|G_{1,T}(t)|$.}}
    \label{fig:cusum-eg}
\end{figure}

As a reminder, our model is
\begin{equation*}
    X_t = \mu + \sigma_t \epsilon_t, ~~ t = 1, \ldots, T,
\end{equation*}
where $\epsilon_t \sim N(0, 1)$, and we assume that $\mu$ is known. Suppose, without loss of generality, that $\mu = 0$ (if this is not the case, we can just replace $X_t$ with $X_t - \mu$). Then a change in variance in $\boldsymbol{X}$ will correspond to a change in mean in $\boldsymbol{X}^2$. Therefore, one approach to detecting a change in variance in $\boldsymbol{X}$ is to look for a change in mean in $\boldsymbol{Y} = \boldsymbol{X}^2$, for example by maximizing the cumulative sum (CUSUM) statistic, which is defined on a interval $(s, e) = \{t : s \leq t \leq e\}$ as
\begin{equation*}
    G_{s,e}(t) = \left(\frac{(t - s + 1) (e - t)}{e - s + 1}\right)^{1/2} \left( \frac{1}{t - s + 1} \sum_{j=s}^t Y_j - \frac{1}{e - t}\sum_{j=t + 1}^e Y_j \right), ~~ t = s, \ldots, e - 1.
\end{equation*}
Figure \ref{fig:cusum-eg} illustrates this process for a single changepoint. We will focus on using the CUSUM statistic with binary segmentation \citep{scott1974scott}; our approach to calculating $p$-values will also extend to variants of this such as wild binary segmentation, as well as to the iterated cumulative sums of squares method of \citep{inclan1994use}, which uses a different statistic based on the cumulative sums of squares.

To detect whether there is a change in mean in an interval $(s, e)$, where $1 \leq s < e \leq T$, we take
\begin{equation*}
    \hat{\tau} = \argmax_t |G_{s,e}(t)|,
\end{equation*}
and determine that there is a changepoint at $\hat{\tau}$ if $|G_{s,e}(\hat{\tau})|$ is greater than some threshold. To detect multiple changepoints using binary segmentation, we start by maximizing $|G_{1,T}(t)|$ across the whole data set; we then split the data at $\hat{\tau}$ and search for changepoints using $G_{s,e}(t)$ calculated on each segment. This process is repeated until no more changepoints are detected.

Using this approach to detecting changes in variance is not always optimal, as the CUSUM statistic is derived under the assumption that the data follows a normal distribution with constant variance, which is not the case for our $\boldsymbol{Y}$: when the variance of $\boldsymbol{X}$ changes, the variance of $\boldsymbol{Y}$ will also change. This leads to a higher false discovery rate in regions of higher variance, and a tendency to miss changepoints in less variable regions. However, it can still work well when  the changes in variance are not too large. It also gives us the advantage of being able to adapt existing post-selection inference methods for the change in mean model, as we shall see in the next section.

\subsubsection{Calculating the post-selection $p$-value} \label{sec:cs-inf}

To calculate the $p$-value in \eqref{eq:pval-con}, we must compute $\mathcal{S} = \{\phi : \hat{\tau} \in \mathcal{M}(\boldsymbol{X}'(\phi))\}$ -- in other words, we need to find the set of $\phi$ values such that our changepoint algorithm detects a changepoint at $\hat{\tau}$. $\mathcal{S}$ is therefore dependent on our choice of changepoint algorithm.
We will focus initially on binary segmentation; the method generalizes straightforwardly to other related algorithms.

To calculate $\mathcal{S}$ we can borrow ideas from methods used for the change in mean model. \cite{hyun2021post} showed that by conditioning on the set of changes $\mathcal{M}(\boldsymbol{X})$, the order in which the changes are found, and the direction of each change (positive or negative), we can write down a set of inequalities which must be satisfied by the CUSUM statistic. These are derived by considering each iteration of binary segmentation in turn, which is possible as conditioning on the order in which changes are found means we know the segments being tested at each iteration. For each iteration where we detect a change, the CUSUM statistics must be largest at the detected value and be above the threshold, while if no change is detected all the CUSUM statistics are below the threshold. We can write out inequalities for each of these possibilities.

For a given segment $(s, e)$, in order to detect a change point at $\hat{\tau}$ we must have $|G_{s,e}(\hat{\tau})| \geq \lambda$, where $\lambda$ is the changepoint detection threshold, and
\begin{equation*}
    \begin{cases}
        (-\hat{d}) \cdot G_{s,e}(\hat{\tau}) > G_{s,e}(t) \\
        (-\hat{d}) \cdot G_{s,e}(\hat{\tau}) > - G_{s,e}(t)
    \end{cases}
    \text{for } t = s, \ldots, e - 1; t \neq \hat{\tau},
\end{equation*}
where $\hat{d} \in \{+1, -1\}$ is the direction of the change. For no changepoint to be detected in the interval $(s, e)$, we must have
\begin{equation*}
    \begin{cases}
        G_{s,e}(t) < \lambda \\
        G_{s,e}(t) > -\lambda 
    \end{cases}
    \text{for } t = s, \ldots, e - 1
\end{equation*}
Hence, we can write down a set of linear inequalities in the data for each iteration of binary segmentation, which must all be satisfied for the particular set of changepoints, with the order and directions of change, to be found. For the change in mean model, the test statistic is defined as $\phi = \boldsymbol{\nu}^\top \boldsymbol{X}$, where
\begin{equation*}
    \nu_t = \begin{cases}
        0 & t \leq \hat{\tau} - h \\
        \frac{1}{h} & \hat{\tau} - h + 1 \leq t \leq \hat{\tau} \\
        -\frac{1}{h} & \hat{\tau} + 1 \leq t \leq \hat{\tau} + h \\
        0 & t \geq \hat{\tau} + h + 1.
    \end{cases}
\end{equation*}
As for our model, we can then introduce perturbations of the data that correspond to changes in $\phi$ whilst fixing the information orthogonal to $\phi$:
\begin{equation*}
    \boldsymbol{X}_{mean}'(\phi) = \boldsymbol{X}_{obs} - \frac{\boldsymbol{\nu}}{||\boldsymbol{\nu}||_2^2} \phi_{obs} + \frac{\boldsymbol{\nu}}{||\boldsymbol{\nu}||_2^2} \phi.
\end{equation*}
Since $G_{s,e}(t)$ is a linear function of $\boldsymbol{X}$, and $\boldsymbol{X}_{mean}'(\phi)$ is a linear function of $\phi$, the CUSUM statistic in terms of $\phi$ is also linear in $\phi$. Hence, we can write the set of inequalities in $G_{s,e}(t)$ as linear inequalities in $\phi$, and they will therefore hold for an interval of $\phi$ values, which it is straightforward to calculate. \cite{jewell2022testing} shows that we can increase power by extending this idea to avoid conditioning on the order and directions of the changes, by calculating the intervals associated with each possible combination of changes which yield $\hat{\tau} \in \mathcal{M}(\boldsymbol{X})$. This gives us the set $\mathcal{S}$ which is the union of all such intervals.

For the change in variance model, we can use the same approach, as the inequalities we write down in terms of $G_{s,e}(t)$ are the same. 
As a reminder, the CUSUM statistic is
\begin{equation*}
    \begin{split}
        G_{s,e}(t) & = \left( \frac{(t - s + 1)(e - t)}{e - s + 1} \right)^{1/2} \left( \frac{1}{t - s + 1} \sum_{j=s}^t Y_j - \frac{1}{e - t}\sum_{j=t + 1}^e Y_j \right) \\
            & = \left( \frac{(t - s + 1)(e - t)}{e - s + 1} \right)^{1/2} \left( \frac{1}{t - s + 1} \sum_{j=s}^t X_j^2 - \frac{1}{e - t} \sum_{j=t + 1}^e X_j^2 \right),
    \end{split}
\end{equation*}
so
\begin{equation}
  \label{eq:cusum-phi}
    G_{s,e}(\phi, t) = \left( \frac{(t - s + 1)(e - t)}{e - s + 1} \right)^{1/2} \left( \frac{1}{t - s + 1} \sum_{j=s}^t X_j'(\phi)^2 - \frac{1}{e - t} \sum_{j=t + 1}^e X_j'(\phi)^2 \right) 
\end{equation}
Using the definition of $\boldsymbol{X}'(\phi)$ from \eqref{eq:x-prime}, we can write ${X_j'}^2(\phi) = a_j + b_j \phi$ where
\begin{equation*}
    a_j = \begin{cases}
        X_j^2 & j \leq \hat{\tau} - h \text{ or } j > \hat{\tau} + h \\
        0 & \hat{\tau} - h < j \leq \hat{\tau} \\
        \frac{1}{1 - \phi_{obs}} X_j^2 & \hat{\tau} < j \leq \hat{\tau} + h
    \end{cases}
\end{equation*}
\begin{equation*}
    b_j = \begin{cases}
        0 & j \leq \hat{\tau} - h \text{ or } j > \hat{\tau} + h \\
        \frac{1}{\phi_{obs}} X_j^2 & \hat{\tau} - h < j \leq \hat{\tau} \\
        - \frac{1}{1 - \phi_{obs}} X_j^2 & \hat{\tau} < j \leq \hat{\tau} + h
    \end{cases}
\end{equation*}
Hence, letting $G_0 = \left( \frac{(t - s + 1)(e - t)}{e - s + 1} \right)^{1/2}$, we can write \eqref{eq:cusum-phi} as
\begin{equation*}
    \begin{split}
        G_{s,e}(\phi, t) & = G_0 \left( \frac{1}{t - s + 1} \sum_{j=s}^t (a_j + b_j) \phi - \frac{1}{e - t} \sum_{j=t + 1}^e (a_j + b_j) \phi \right) \\
            & = G_0 \left( \left( \frac{1}{t - s + 1} \sum_{j=s}^t a_j - \frac{1}{e - t} \sum_{j=t + 1}^e a_j \right) + \left( \frac{1}{t - s + 1} \sum_{j=s}^t b_j - \frac{1}{e - t} \sum_{j=t + 1}^e b_j \right) \phi \right).
    \end{split}
\end{equation*}
This is linear in $\phi$, so we can once again compute each possible interval by resolving a set of linear inequalities in $\phi$. We can use the method of \cite{jewell2022testing} to efficiently compute $\mathcal{S}$ (and hence $p$).

\subsection{Likelihood-ratio based detection of changes} \label{sec:lrs}

\begin{figure}
    \centering
    \includegraphics[width=0.98\linewidth]{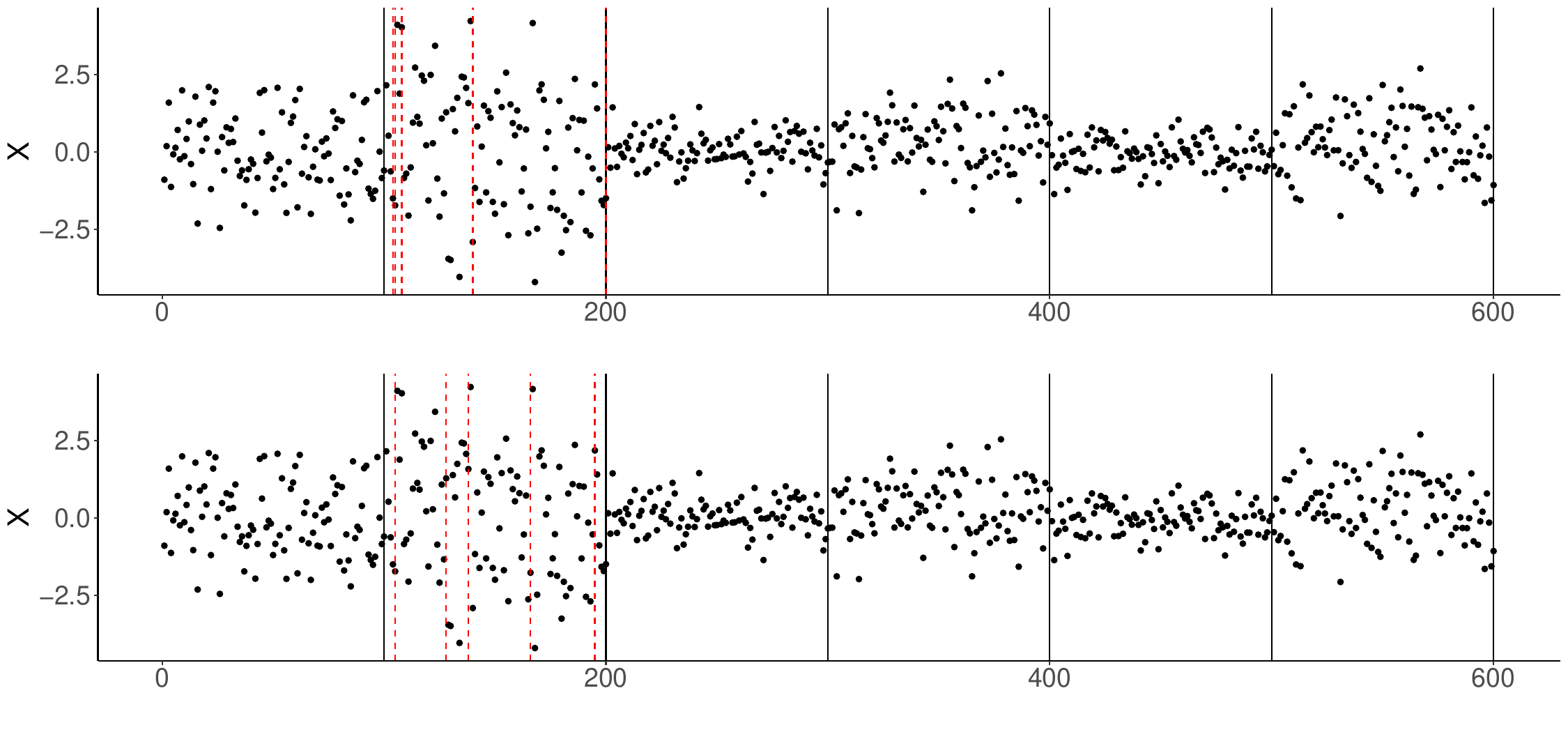}
    \caption{\small{Estimating changepoints using binary segmentation and wild binary segmentation with the CUSUM statistic for a simulated data set with multiple changes. The solid black lines correspond to true changes, the dashed red lines to estimated changepoints. The method tends to over-estimate changepoints within the region of highest variance, whilst missing changes in other regions.}}
    \label{fig:cs-vs-lrs-1}
\end{figure}

Whilst post-selection inference is relatively straightforward for the piecewise constant variance model when estimating changepoint locations using the CUSUM statistic, the use of the CUSUM statistic can lead to poor detection of changes. For example, Figure \ref{fig:cs-vs-lrs-1} shows an example of a simulated data set with 5 changepoints, where we apply binary segmentation and wild binary segmentation to detect changepoints. Both methods overestimate the number of changes in the region with highest variance, but fail to detect changes in other regions.

We can achieve more accurate changepoint detection by maximizing the likelihood ratio statistic, which on an interval $(s, e)$ is defined as
\begin{equation}
    \Lambda_{s,e} (\tau) = (s - e + 1) \log \sum_{t=s}^e X_t^2 - (\tau - s + 1) \log \sum_{t=s}^{\tau} X_t^2 - (e - \tau) \log \sum_{t=\tau + 1}^e X_t^2
  \label{eq:lrs}
\end{equation}
To estimate a single changepoint, we calculate $\Lambda_{1,T}(\tau)$ for $\tau = 1, \ldots, T - 1$, and take $\hat{\tau} = \argmax_{\tau} \Lambda_{1,T}(\tau)$. For multiple changepoints, we can either use an iterative method such as binary segmentation, or a penalized likelihood method such as PELT \citep{killick2012optimal}. The approach we develop below will apply in both cases.

\begin{table}
    \centering
    \caption{\small{Proportion of runs in which binary segmentation detected a changepoint within 10 of each true changepoint, using CUSUM and likelihood ratio approaches.}}
    \begin{tabular}{ccc}
        \hline
            &~~~~CUSUM ~~~~& ~~LR Statistic~~ \\
        \hline
            ~~$\tau = 100$~~ & 0.755 & 0.915 \\ 
            $\tau = 200$ & 0.972 & 0.992 \\
            $\tau = 300$ & 0.012 & 0.914 \\
        \hline
    \end{tabular}
    \label{tab:cusum-vs-lrs}
\end{table}

Table \ref{tab:cusum-vs-lrs} demonstrates how this leads to increased accuracy compared with using the CUSUM statistic. We simulated data from a model with $T = 400$ and changepoints at $\tau = 100, 200, 300$; the variances on the four segments being $\{1, 4, 0.25, 1\}$, and for each simulated data set we estimated three changes using binary segmentation with both the CUSUM and likelihood ratio statistics. For each true changepoint location, the table records the proportion of runs in which a changepoint was detected within 10 of that location. We see that CUSUM does reasonably well at detecting the first two changes, but does very badly at locating the third, smaller change. Using the likelihood ratio statistic, however, all three changepoints are detected in over $90\%$ of cases.

\subsubsection{Post-selection inference} \label{sec:lr-psi}

As a reminder, the post-selection $p$-value we want to calculate is 
\begin{equation*}
    \Pr \left( \phi \leq \phi_{lower} \text{ or } \phi \geq \phi_{upper} ~ | ~ \phi \in \mathcal{S} \right),
\end{equation*}
where $\mathcal{S} = \{\phi : \hat{\tau} \in \mathcal{M}(\boldsymbol{X}'(\phi))\}$.
Previously, when detecting changes based on the CUSUM statistic, we were able to write down a set of linear inequalities in $\phi$ which defined the set $\mathcal{S}$. However, while we can write down inequalities in $\Lambda_{s,e}(\phi, \tau)$ as we did for $G_{s,e}(\phi, \tau)$, $\Lambda_{s,e}(\phi, \tau)$ is not a linear function of $\phi$, see the Supplementary Material for more details.

\begin{figure}
    \centering
    \includegraphics[width=0.96\linewidth]{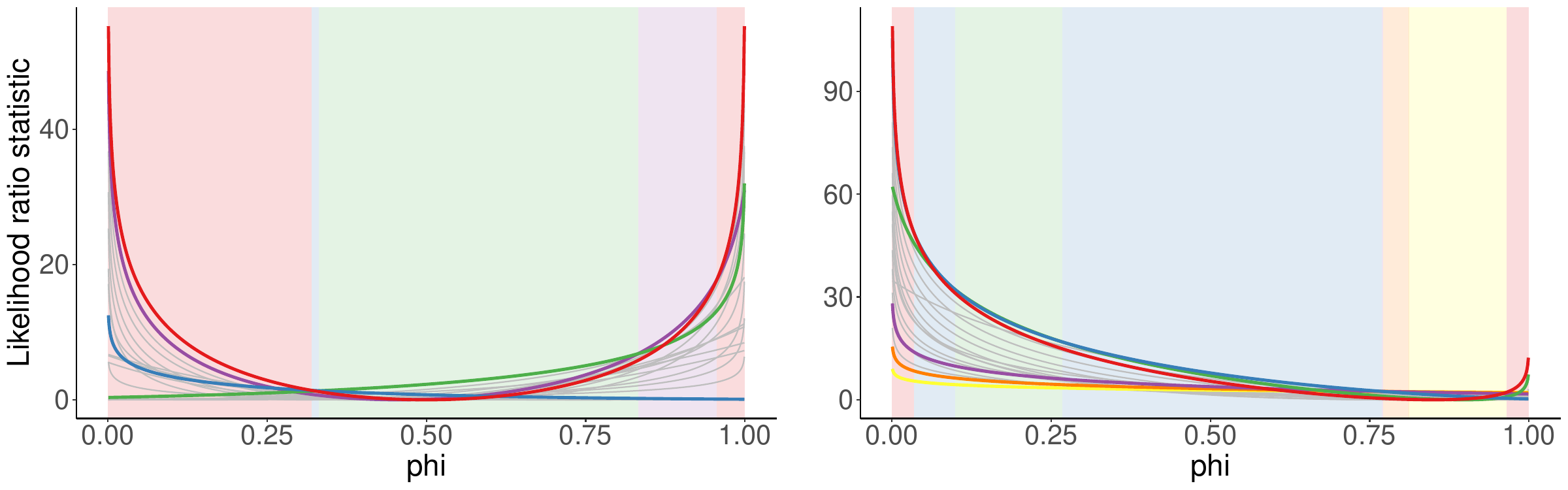}
    \caption{\small{Example plots of $\Lambda$ against $\phi$; each line corresponds to $\Lambda(\tau, \phi)$ for a different value of $\tau \in 1, \ldots, T - 1$. Each shaded region corresponds to a range of $\phi$ values where a different $\tau$ maximizes $\Lambda(\tau, \phi)$.}}
    \label{fig:lrs-plot}
\end{figure}

However $\mathcal{S}$ still consists of a union of intervals as the inequalities are smooth functions of $\phi$.
To illustrate this, Figure \ref{fig:lrs-plot} shows a plot of $\Lambda(\tau, \phi)$ against $\phi$ for a simulated data set for different values of $\tau$; the red line corresponds to $\hat{\tau}$. The values of $\phi$ which are in $\mathcal{S}$ are hence those for which the red line is higher than any of the other lines on the plot.

Although we cannot compute $\mathcal{S}$ directly, we can find out whether a given $\phi$ is in $\mathcal{S}$ by computing $\boldsymbol{X}'(\phi)$, as in \eqref{eq:x-prime}, and re-running the changepoint algorithm to obtain the set of estimated changepoints $\mathcal{M}(\boldsymbol{X}'(\phi))$. Hence, one way to approximate $\mathcal{S}$ is by randomly sampling $\phi$ values and calculating whether each is in $\mathcal{S}$. We can then estimate $p$ by calculating what proportion of $\phi$ values which are in $\mathcal{S}$ are also in the critical region $\mathcal{C} = \{\phi : \phi \leq \phi_{lower} \text{ or } \phi \geq \phi_{upper} \}$:
\begin{equation}
  \label{eq:p-hat}
    \hat{p} = \frac{\sum_{i=1}^N \mathbb{I}_{\{\hat{\tau} \in \mathcal{M}(\boldsymbol{X}'(\phi^{(i)})), ~~ \phi^{(i)} \in \mathcal{C} \}}}{\sum_{i=1}^N \mathbb{I}_{\{\hat{\tau} \in \mathcal{M}(\boldsymbol{X}'(\phi^{(i)}))\}}}.
\end{equation}

However, sampling $\phi$ from its unconditional distribution $\text{Beta} \left( \frac{h}{2}, \frac{h}{2} \right)$ often leads to very slow convergence of $\hat{p}$, as $\Pr(\phi \in \mathcal{S})$ is typically small, so only a small proportion of samples will contribute to $\hat{p}$: the rest are discarded.

Since sampling in this way requires re-applying the changepoint algorithm for each $\phi^{(i)}$, which may be computationally demanding (the length of time taken to apply the algorithm will scale with the number of data points and the number of detected changes), using a large number of samples to estimate the $p$-value is undesirable. We therefore want to find ways to reduce the number of samples necessary to get a good $p$-value estimate $\hat{p}$. 

One way to reduce the variance of $\hat{p}$, without increasing the computation time, is to use stratified sampling. For $i = 1, \ldots, N$, we sample $z^{(i)} \sim U(\frac{i-1}{N}, \frac{i}{N})$. We then calculate $\phi^{(i)}$ to satisfy $\Pr (\phi < \phi^{(i)}) = z^{(i)}$ (see Section \ref{sec:strat} in the Supplementary Material). However, it is often the case that $\Pr(\phi \in \mathcal{S})$ is of the order of $\frac{1}{N}$ or smaller, in which case using stratified sampling is unlikely to improve the estimate $\hat{p}$, as we will still have very few $\phi^{(i)} \in \mathcal{S}$. To overcome this we resort to using simulation to estimate the intervals that contribute to $\mathcal{S}$.

\subsubsection{Estimating $\mathcal{S}$ using Gaussian processes} \label{sec:gp}

If we can calculate $\mathcal{S}$ then it is straightforward to calculate the $p$-value. For any value of $\phi$ we can represent our belief about $\mathbb{I}_{\phi \in \mathcal{S}}$ by a probability $p(\phi) = \Pr(\mathbb{I}_{\phi \in \mathcal{S}})$. Our approach is to model the function $p(\phi)$ as a Gaussian process (GP) \cite[see e.g.][]{williams2006gaussian}, and update our beliefs about this function by evaluating $\mathbb{I}_{\phi \in \mathcal{S}}$ at a set of suitably chosen values for $\phi$. Whilst a GP does not enforce that $p(\phi)$ lies in $[0,1]$, our implementation below ensures that the mean of the GP does lie in this interval.

To implement this, we first need to choose a kernel function, $\boldsymbol{K}(\phi, \phi')$, which specifies the covariance between each pair of $\phi$ values. We choose an exponential kernel $\boldsymbol{K}(\phi_1, \phi_2) = e^{-\frac{1}{2l^2} |\phi_1 - \phi_2|}$, which is Markov. (This requires us to choose the value of $l$: we investigate this in Section \ref{sec:finding-l} of the Supplementary Material, where we find that at long as $l$ is at least 1, the value we use does not have a noticeable effect on the results.) If we have evaluated $\mathbb{I}_{\phi \in \mathcal{S}}$ at $\phi_1 < \cdots < \phi_N$, and $\phi_i < \phi < \phi_{i+1}$ the Markov property means that our Gaussian approximation to $p(\phi)$ will only depend on the values of $\mathbb{I}_{\phi_i \in \mathcal{S}}$ and $\mathbb{I}_{\phi_{i+1} \in \mathcal{S}}$. Furthermore, for such a kernel the mean of the Gaussian process will be an average of these values and the prior mean -- and thus will always lie within $[0,1]$.

We first choose a stratified sample from $U(0, 1)$, $\phi_1 < \cdots < \phi_N$, and calculate $\mathbb{I}_{\phi_i \in \mathcal{S}}$ for each $i = 1, \ldots, N$. 
As before, we have to re-apply the changepoint algorithm to calculate each $\mathbb{I}_{\phi_i \in \mathcal{S}}$.

We can then calculate the estimate of $p(\phi)$ at discrete locations $\boldsymbol{\phi}' = \{\phi_1', \ldots, \phi_N'\}$, as the mean of our GP:
\begin{equation*}
    \hat{\boldsymbol{p}}(\boldsymbol{\phi}') = \boldsymbol{K}(\boldsymbol{\phi}, \boldsymbol{\phi}')^\top \boldsymbol{K}(\boldsymbol{\phi}, \boldsymbol{\phi})^{-1} \boldsymbol{p}(\boldsymbol{\phi}),
\end{equation*}
where we use $\boldsymbol{K}(\boldsymbol{\phi}, \boldsymbol{\phi}')$ to denote the matrix whose $(i,j)$th entry is $K(\phi_i,\phi'_j)$. The most expensive step here computationally is calculating the inverse of the $N\times N$ matrix $\boldsymbol{K}(\boldsymbol{\phi}, \boldsymbol{\phi})$, but the Markov property of our kernel means that this scales linearly with $N$.

Let $\pi(\phi)$ denote the unconditional distribution of $\phi$, i.e. $Beta(\frac{h}{2}, \frac{h}{2})$. Let $q(\phi)$ denote the conditional distribution of $\phi | \phi \in \mathcal{S}$:
\begin{equation*}
  \begin{split}
      q(\phi) & \propto \pi(\phi) \cdot \mathbb{I}_{\phi \in \mathcal{S}}
  \end{split}
\end{equation*}
We can estimate this by replacing $\mathbb{I}_{\phi \in \mathcal{S}}$ with the mean of this quantity under our GP model. This is just the mean of the GP, and gives the estimate, $\hat{q}(\phi)$, where
\begin{equation*}
    \hat{q}(\phi) = \frac{\hat{p}(\phi) \pi(\phi)}{\int_0^1 \hat{p}(\phi) \pi(\phi) d\phi},
\end{equation*}
which is easy to evaluate numerically.

To estimate the $p$-value, we have two options. The $p$-value is
\begin{equation}
    p = \Pr (\phi \leq \phi_{lower} \text{ or } \phi \geq \phi_{upper} ~|~ \phi \in \mathcal{S}) = q(\phi \leq \phi_{lower} \text{ or } \phi \geq \phi_{upper}).
  \label{eq:pos-p}
\end{equation}
Since $\hat{q}(\phi)$ is an estimate of the density of $\phi | \phi \in \mathcal{S}$, we can use this directly to estimate the $p$-value, i.e.
\begin{equation*}
    \hat{p} = \hat{q}(\phi \leq \phi_{lower} \text{ or } \phi \geq \phi_{upper}).
\end{equation*}
Alternatively, we can use importance sampling with $\hat{q}$ as the proposal distribution. This means that we sample $\tilde{\phi}_1, \ldots, \tilde{\phi}_{\tilde{N}} \sim \hat{q}(\phi)$; then
\begin{equation}
    \hat{p} = \frac{ \sum_{j=1}^{\tilde{N}} \mathbb{I} \left( \tilde{\phi}_j \in \mathcal{S} ~ \text{and} ~ \left( \tilde{\phi}_j \leq \phi_{lower} \text{ or } \tilde{\phi}_j \geq \phi_{upper} \right) \right) \frac{\pi(\tilde{\phi}_j)}{\hat{q}(\tilde{\phi}_j)} } { \sum_{j=1}^{\tilde{N}} \mathbb{I} \left( \tilde{\phi}_j \in \mathcal{S} \right) \frac{\pi(\tilde{\phi}_j)}{\hat{q}(\tilde{\phi}_j)} }.
  \label{eq:pos-imp-p}
\end{equation}
Since we have already calculated $\mathbb{I}_{\phi \in \mathcal{S}}$ for the $\phi$ used to fit the GP, we can also include these points in our importance sampling estimator.

\begin{figure}
    \centering
    \includegraphics[width=0.98\linewidth]{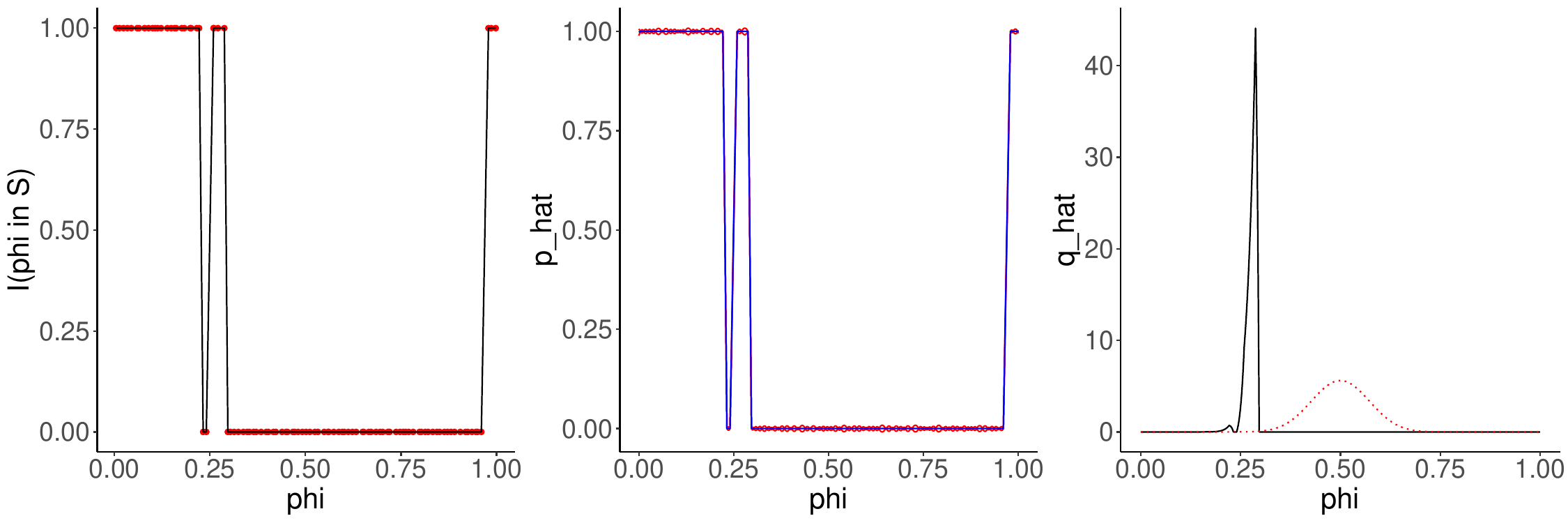}
        \put(-300,-8){\footnotesize{(a)}}
        \put(-180,-8){\footnotesize{(b)}}
        \put(-60,-8){\footnotesize{(c)}}
    \caption{\small{Estimating $q(\phi)$ using a Gaussian process. Left: $\mathbb{I}_{\phi \in \mathcal{S}}$ plotted against $\phi$ for 100 samples. Middle: posterior mean of the Gaussian process (blue line) with $95\%$ predicted interval (red) for $\mathbb{I}_{\phi \in \mathcal{S}}$. Right: Estimate of $\hat{q}(\phi)$ (black line); the dotted red line shows $\pi(\phi)$.}}
    \label{fig:gp-fig}
\end{figure}

Figure \ref{fig:gp-fig} shows an outline of this process. First, we sample $\phi_1, \ldots, \phi_N$ --- we here sample from $U(0, 1)$ to ensure that we get good coverage of the range of $\phi$ values. For each $\phi_i$, we calculate $\boldsymbol{X}'(\phi_i)$ and apply the changepoint algorithm to evaluate $p(\phi_i) = \mathbb{I}_{\phi_i \in \mathcal{S}}$. Panel (a) shows a plot of $p(\phi)$ for our samples. Then we fit a Gaussian process: panel (b) shows the posterior mean $\hat{p}(\phi)$. We then estimate $\hat{q}(\phi)$ as described above: this is shown by the black line in panel (c), with the dotted red line showing $\pi(\phi)$ for comparison.

\section{Extensions} \label{sec:ext}

\subsection{Different null hypotheses} \label{sec:ext-1}

There are alternative choices for the null hypothesis we may want to test, in particular to deal with the possibility of detected changepoints being close together. For example, we could decide to truncate the region if we find another changepoint within $h$ of the changepoint being tested. Alternatively, if we denote our changepoint of interest as $\hat{\tau}_j$, we can take the null hypothesis that there is no changepoint within $\{\hat{\tau}_{j-1} + 1, \ldots, \hat{\tau}_{j+1} - 1\}$ (by convention setting $\hat{\tau}_0 = 0$ and $\hat{\tau}_{K+1} = T$, where $K$ is the number of changepoints), or define the region we test to that closer to $\hat{\tau}_j$ than any other detected changepoint \cite[see][]{jewell2022testing,carrington2023improving}.

In general, the methods we have described above will work in the same way for these different alternatives. However, in the case where the region of interest is determined by the locations of other estimated changepoints, it is necessary to redefine the conditioning set to take account of this. If the window is determined by the locations of neighbouring changepoints, then we also need to condition on the fact that these changepoint locations are included in the set of changepoints $\mathcal{M}(\boldsymbol{X})$. This can be done, for example, by conditioning on the set of all estimated changepoints --- i.e. $\mathcal{M}(\boldsymbol{X}'(\phi)) = \mathcal{M}(\boldsymbol{X}_{obs})$ --- rather than just on the fact that $\hat{\tau} \in \mathcal{M}(\boldsymbol{X}'(\phi))$. To do this, for the CUSUM method we construct the set of intervals in the same way, but only choose the ones for which this new condition holds to be in $\mathcal{S}$. For the more general case, it is straightforward to implement this.

\subsection{Different changepoint detection methods} \label{sec:ext-2}

We have focused on two approaches to post-selection inference for the change in variance model. The CUSUM approach (Section \ref{sec:cusum}) depends on being able to write down a set of linear inequalities in $\phi$ which can be resolved to obtain the set $\mathcal{S}$. We have focused on doing this for binary segmentation; however, it can be easily shown that we can do the same for variants -- such as wild binary segmentation -- which also involve maximizing $G_{s,e}(t)$ on a series of intervals. 
For our second approach (Section \ref{sec:lrs}), we can use any method for changepoint estimation, since we just re-apply the algorithm for each sampled $\phi$.

For both methods, we will generate data sets for different values of $\phi$ and re-run our changepoint detection method on each data set. This is the main computational cost of both methods, though they differ in terms of the number of new data sets that have to be analysed. In the former method this is equal to the number of intervals of $\phi$ that correspond to a different solution path of e.g. binary segmentation. In most examples we have considered this is fewer than 10. For the second method the number of data sets depends on the Monte Carlo accuracy needed when estimating the $p$-value but tends to be of the order of 100s.

Often one can reduce the computational cost through early stopping when we run the changepoint detection algorithm. If we are conditioning on $\hat{\tau}\in\mathcal{M}$ then we can stop once this condition holds. By comparison if we are conditioning on $\mathcal{M}$ we can stop if ever we detect a change not in $\mathcal{M}$ for our new data. Also we can often re-use calculations. For example, to implement methods such as wild binary segmentation and narrowest over threshold we have to calculate the likelihood ratio statistic on a set of intervals $[s, e] \subseteq [1, T]$. The set of intervals is the same each time we implement the algorithm on $\boldsymbol{X}'(\phi)$. Since $\Lambda_{s,e}(t, \phi)$ is constant on intervals which do not overlap with the region of interest (i.e., $e \leq \hat{\tau} - h$ or $s \geq \hat{\tau} + h + 1$), for these intervals we only have to calculate $\Lambda_{s,e}(t)$ once. Therefore, at each sampling iteration we only have to re-calculate $\Lambda_{s,e}(t, \phi)$ on intervals which overlap with the region of interest.

\subsection{Increasing power by conditioning on less information} \label{sec:ext-3}

So far, we have calculated $p$-values by conditioning on all aspects of the data except for the test statistic $\phi$. However, whilst it is necessary to condition on sufficient statistics for the parameters which are not specified under $H_0$, it is possible to define a $p$-value which is not conditional on $\boldsymbol{W} = \{W_1^l, \ldots, W_{h-1}^l, W_1^r, \ldots, W_{h-1}^r\}$, i.e.
\begin{equation*}
    \Pr \left( \phi \leq \phi_{lower} \text{ or } \phi \geq \phi_{upper} ~ | ~ \hat{\tau} \in \mathcal{M}(\boldsymbol{X}'(\phi, \boldsymbol{W})) \right).
\end{equation*}
where here both $\phi$ and $\boldsymbol{W}$ are allowed to vary.

This adds complication in calculating the conditional distribution of $\phi$ as the $W$'s are nuisance parameters: although, unconditionally, $\boldsymbol{W}$ and $\phi$ are independent, when the distribution of $\phi$ is calculated truncated to the region of interest this is not the case, as the truncation region also depends on $\boldsymbol{W}$. Conditioning on the observed value of $\boldsymbol{W}$ is one way of dealing with this. An alternative approach, as in \cite{carrington2023improving} for the change in mean model, is to sample values of $\boldsymbol{W}$, calculate the truncation region for each, and then to calculate the $p$-value as a weighted average of these individual $p$-values, i.e.
\begin{equation*}
    \hat{p} = \frac{\sum_{j=1}^{N_W} \Pr \left(\phi \geq \phi_{upper} \text{ or } \phi \leq \phi_{lower}, \hat{\tau} \in \mathcal{M}(\boldsymbol{X}), ~|~ \boldsymbol{W} = \boldsymbol{W}^{(j)}\right)}{\sum_{j=1}^{N_W} \Pr \left(\hat{\tau} \in \mathcal{M}(\boldsymbol{X}) ~|~ \boldsymbol{W} = \boldsymbol{W}^{(j)}\right)}.
\end{equation*}
Under $H_0$, $W_i^l, W_i^r \sim \text{Beta} \left( \frac{i}{2}, \frac{1}{2} \right)$. As long as we include the observed value of $\boldsymbol{W}$ as one of the samples, the $p$-value estimates we obtain will be valid, in the sense that they will follow a uniform distribution under $H_0$. We investigate applying this method on simulated data in Section \ref{sec:sims-increasing-power}.

\section{Simulations} \label{sec:sims}

In this section we implement our methods on simulated data. All simulation results are based on 1000 replicates.

\subsection{Detecting changes using CUSUM} \label{sec:sims-cs}

\begin{figure}
    \centering
    \includegraphics[width=\linewidth]{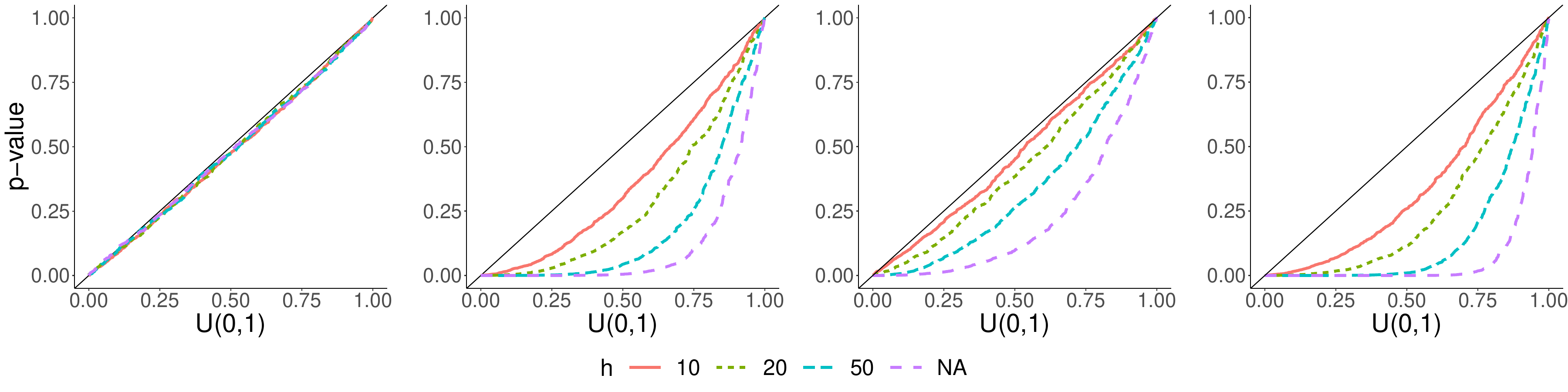}
        \put(-330, 95){\footnotesize{$\sigma_2^2 = 1$}}
        \put(-320, -5){\footnotesize{(a)}}
        \put(-240, 95){\footnotesize{$\sigma_2^2 = 0.25$}}
        \put(-230, -5){\footnotesize{(b)}}
        \put(-142, 95){\footnotesize{$\sigma_2^2 = 2$}}
        \put(-132, -5){\footnotesize{(c)}}
        \put(-50, 95){\footnotesize{$\sigma_2^2 = 4$}}
        \put(-42, -5){\footnotesize{(d)}}
    \caption{\small{QQ plots of $p$-values obtained using the CUSUM method. We set $T = 200$, and sampled data under $H_0$ (no change) and with a single changepoint sampled randomly from $1, \ldots, T$. The variance before the change was $\sigma_1^2 = 1$ in each case; the variance after the change is $\sigma^2 \in \{1, 0.25, 2, 4\}$. In each case we estimate a single changepoint using binary segmentation. We calculated $p$-values for different values of $h$, including the case where the window size is equal to the whole data set.}}
    \label{fig:cusum}
\end{figure}

Figure \ref{fig:cusum} shows QQ plots of $p$-values obtained by sampling from a model with $T = 200$. In (a) we sample from $H_0$; in (b)--(d) we sample from a model with a single change at $\tau = 100$, where in each case the variance before the changepoint is 1, and the variance after the changepoint is 0.25, 2, 4. We see that under $H_0$, $p \sim U(0, 1)$, and under $H_1$ the test has power to detect changes, which increases with the window size $h$ and the size of the change.

\begin{figure}
    \centering
    \includegraphics[width=0.98\linewidth]{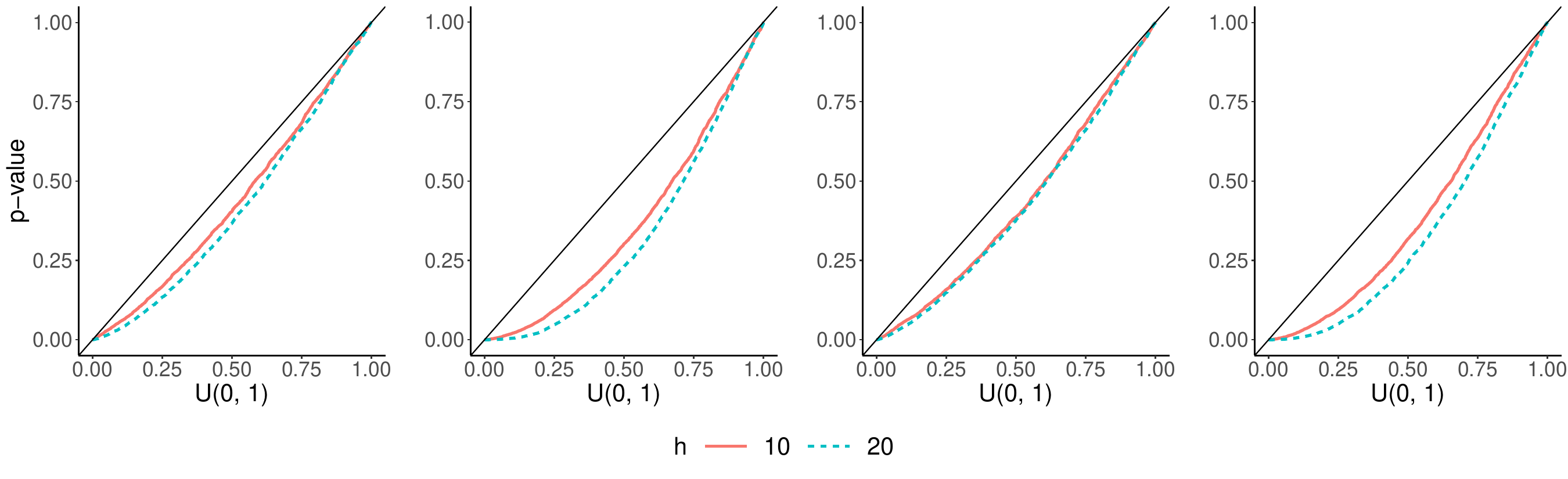}
        \put(-320, 110){\footnotesize{$\sigma_2^2 = 2$}}
        \put(-312, -5){\footnotesize{(a)}}
        \put(-228, 110){\footnotesize{$\sigma_2^2 = 4$}}
        \put(-220, -5){\footnotesize{(b)}}
        \put(-140, 110){\footnotesize{$\sigma_2^2 = 2$}}
        \put(-130, -5){\footnotesize{(c)}}
        \put(-50, 110){\footnotesize{$\sigma_2^2 = 4$}}
        \put(-42, -5){\footnotesize{(d)}}
    \caption{\small{QQ plots of $p$-values obtained using the CUSUM method, where we sample data from a model with $T = 200$ and 4 changepoints. In (a) and (b), the changes are equally spaced at $\tau = 40, 80, 120, 160$; in (c) and (d) they are random, sampled uniformly from $\{h, \ldots, T - h\}$ with the constraint that the distance between any two changes is at least $h$. The plots show sorted $p$-values for all estimated changepoints (up to 4 for each sampled data set).}}
    \label{fig:cusum-2}
\end{figure}

To see what happens when we have multiple changes, Figure \ref{fig:cusum-2} shows similar plots where we simulate from a model with 4 changes, with $T = 200$ and $h = 10, 20$. In panels (a) and (b) we simulated from a model with 4 equally spaced changes, $\tau = 40, 80, 120, 160$. In (c) and (d), changepoints were chosen uniformly from $(h, T - h)$, with the constraint that the minimum distance between changepoints was at least $h$.

\subsection{Detecting changes using the likelihood-ratio test} \label{sec:sims-lrs}

In Section \ref{sec:gp}, we discussed how we can estimate the $p$-value using a Gaussian process. To implement this we must decide on the number of random samples $N$, and also to decide whether we will estimate $\hat{p}$ directly from the posterior of the Gaussian process, as in \eqref{eq:pos-p}, or use importance sampling from the posterior distribution to estimate $\hat{p}$, as in \eqref{eq:pos-imp-p}.
We here investigate difference choices of this for several different simulated scenarios. In particular, we are interested in knowing what happens when the number of samples $N$ is relatively small.

\begin{figure}
    \includegraphics[width=0.85\linewidth]{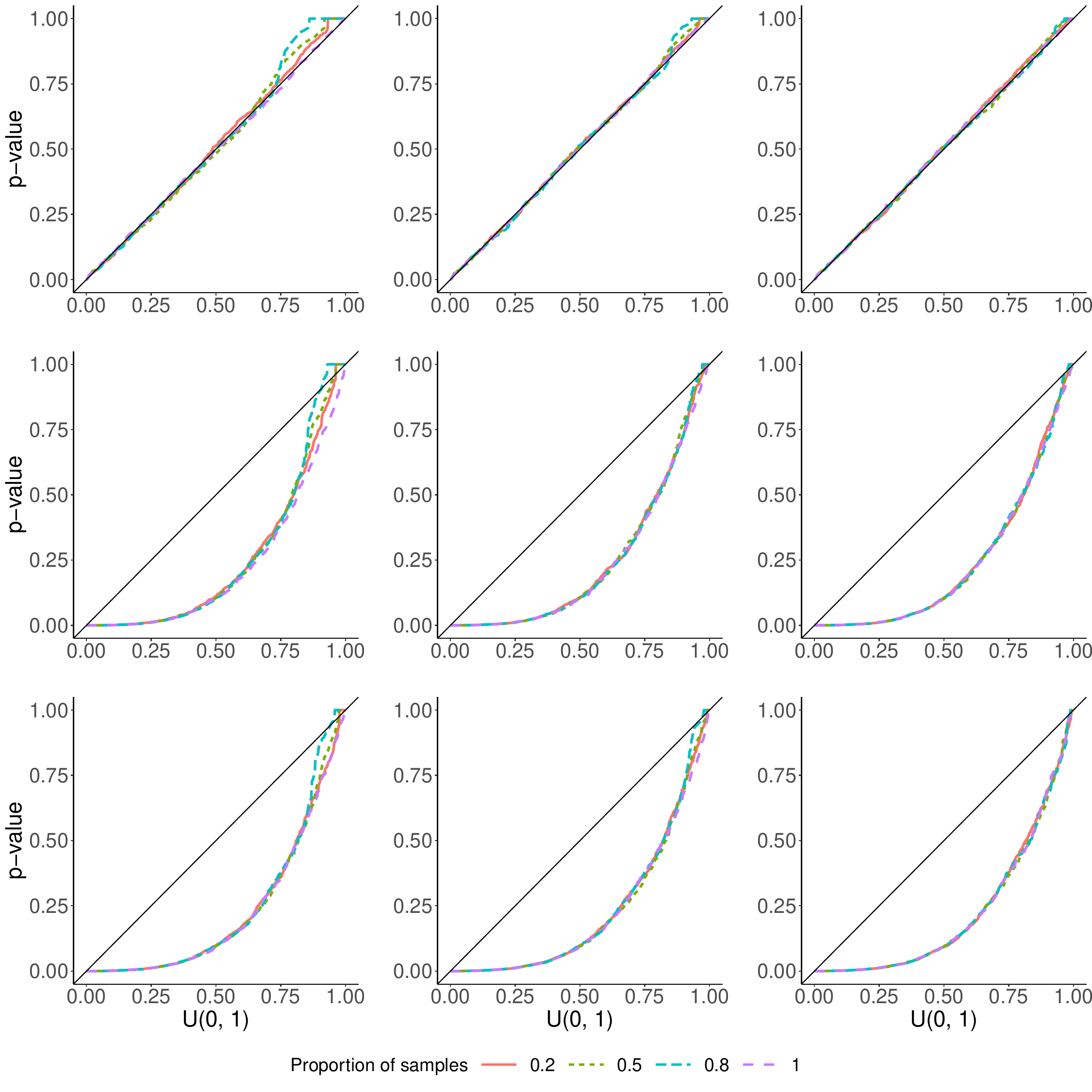}
        \put(-265, 318){\footnotesize{$N = 50$}}
        \put(-158, 318){\footnotesize{$N = 100$}}
        \put(-55, 318){\footnotesize{$N = 200$}}
        \put(0, 270){\footnotesize{$H_0$}}
        \put(0, 180){\footnotesize{$\tau_1 = \frac{T}{4}, \tau_2 = \frac{T}{2}$,}}
        \put(0, 170){\footnotesize{$\tau_3 = \frac{3T}{4}$}}
        \put(0, 80){\footnotesize{$\tau_1, \tau_2, \tau_3$ }}
        \put(0, 70){\footnotesize{random}}
    \caption{\small{QQ plots of $p$-values obtained using the likelihood ratio for three scenarios -- no change, three equally spaced changes, and three random changes -- using binary segmentation to estimate changepoints. Each row of the figure corresponds to a scenario, and columns correspond to the number of samples used. Each line corresponds to the proportion of samples that were used to fit the Gaussian process.}}
    \label{fig:is-vs-gp}
\end{figure}

Figure \ref{fig:is-vs-gp} shows QQ plots of estimated $p$-values we obtain on simulated data for three different scenarios: (1) no change; (2) three changes at fixed equidistant locations, $\tau = \frac{T}{4}, \frac{T}{2}, \frac{3T}{4}$; (3) three changes at random locations, sampled uniformly from $\{h, \ldots, T - h\}$ and constrained to be at least $h$ apart. In each case we set $T = 200$ and $h = 20$. For the cases with three changes, the variances on each segment were $(1, 4, 1, 0.25)$. For each simulated data set, we estimated changepoints using binary segmentation (Figure \ref{fig:is-vs-gp}) and calculated an estimate of the $p$-value corresponding to the first detected changepoint using a Gaussian process (as outlined in Section \ref{sec:gp}), and using a Gaussian process followed by importance sampling from the posterior distribution, for $N = 50, 100, 200$.

Under $H_0$, the $p$-values obtained using just a Gaussian process (i.e. the proportion of GP samples is 1) lie along the straight line implying they follow the distribution $U(0, 1)$, whilst if we use a GP followed by importance sampling the $p$-values are slightly conservative for small $N$, although still close to uniform. Similarly, when we sample data with changepoints, for small $N$ we get slightly greater power if we just use a GP. For $N = 200$ there seems to be very little difference between these methods. Overall, it appears that using just a Gaussian process generally gives at least as good results as using a Gaussian process followed by importance sampling.

How do we know when we have enough samples? We see that $N = 50$ seems to be sufficient in the above cases, as the estimated $p$-values follow a uniform distribution under $H_0$, and under $H_1$ increasing $N$ from 50 to 200 does not seem to lead to increased power. It is possible that we would need more samples if the number of data points $T$ or the number of changes was larger. 

\subsection{Increasing power} \label{sec:sims-increasing-power}

\begin{figure}
    \centering
    \includegraphics[width=0.56\linewidth]{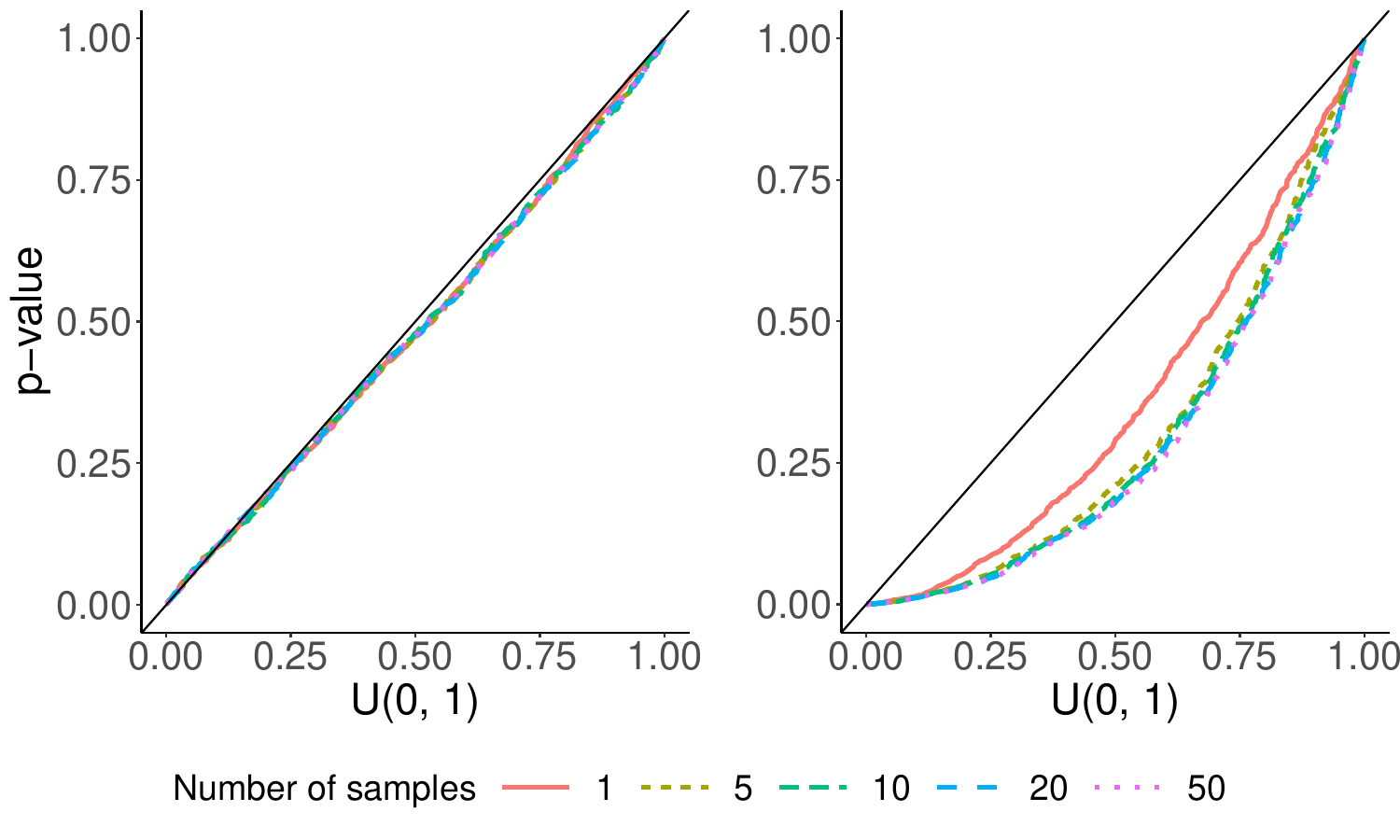}
    \put(-150, 125){\footnotesize{$H_0$}}
    \put(-45, 125){\footnotesize{$H_1$}}
    \caption{\small{QQ plots of $p$-values obtained when estimating a single changepoint using the CUSUM statistic, and calculating $p$-values using the method of \ref{sec:ext-3}.}}
    \label{fig:increasing-power-cusum}
\end{figure}

Figure \ref{fig:increasing-power-cusum} shows $p$-values we obtained where we estimated a single changepoint using the CUSUM statistic and used the method described in Section \ref{sec:ext-3} to calculate $p$-values. We can see that under $H_0$ (left) we obtain valid $p$-values for any value of the number of samples $N_W$. When we sample from a model with a single change at $\frac{T}{2}$ (right), increasing $N_W$ from 1 to 5 gives greater power against $H_0$, although increasing it beyond this does not seem to make much difference.

\begin{figure}
    \centering
    \includegraphics[width=0.98\linewidth]{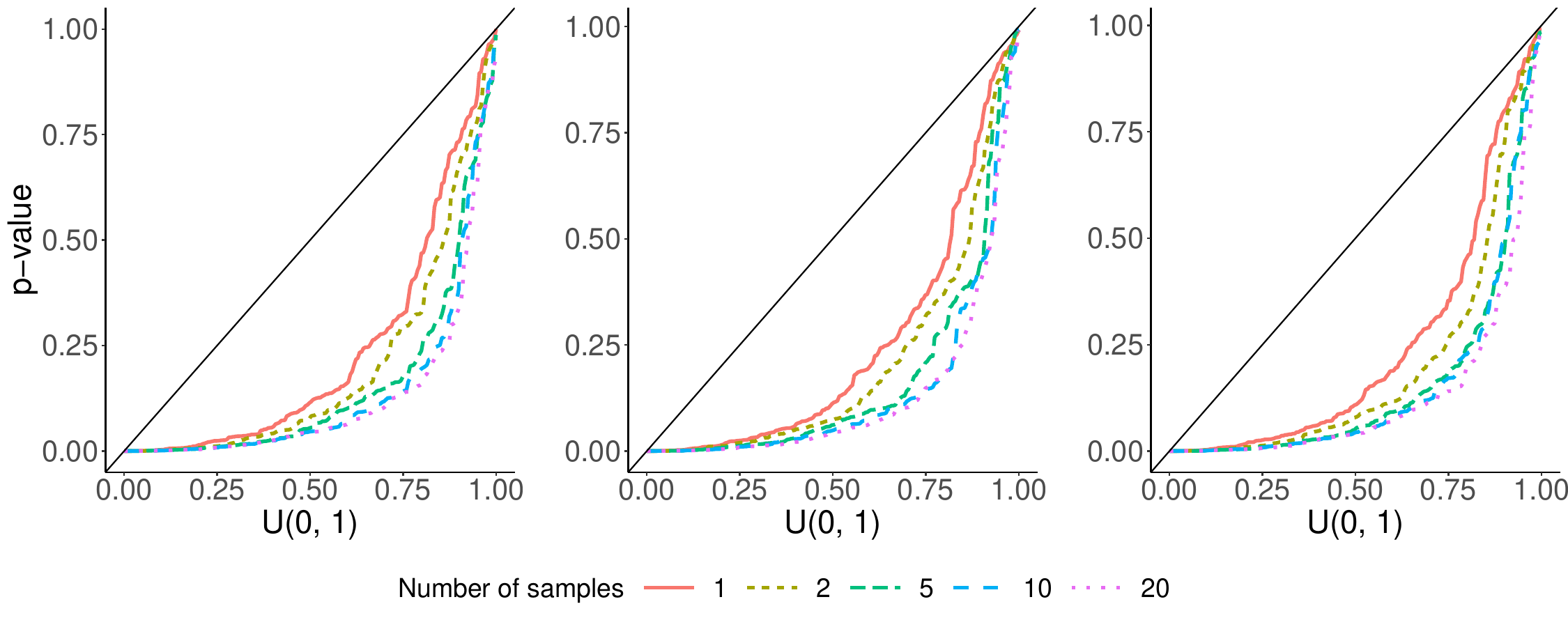}
        \put(-305, 145){\footnotesize{$N_{\phi} = 50$}}
        \put(-190, 145){\footnotesize{$N_{\phi} = 100$}}
        \put(-65, 145){\footnotesize{$N_{\phi} = 200$}} 
        \caption{$P$-values obtained sampling under $H_1$, and estimating a single changepoint using a Gaussian process, using the method of \ref{sec:ext-3}. $N_{\phi}$ is the number of $\phi$ samples for the Gaussian process.}
    \label{fig:increasing-power-lrs}
\end{figure}

Figure \ref{fig:increasing-power-lrs} shows similar plots where we simulate from a model with a change at $\frac{T}{2}$, and use the likelihood ratio to estimate changepoints, and a Gaussian process with $N_{\phi}$ samples to estimate $p$-values. We see a similar pattern, where increasing the number of $W$ samples initially leads to greater power. 

\section{Application to financial data} \label{sec:real}

\begin{figure}
    \centering
    \includegraphics[width=\linewidth]{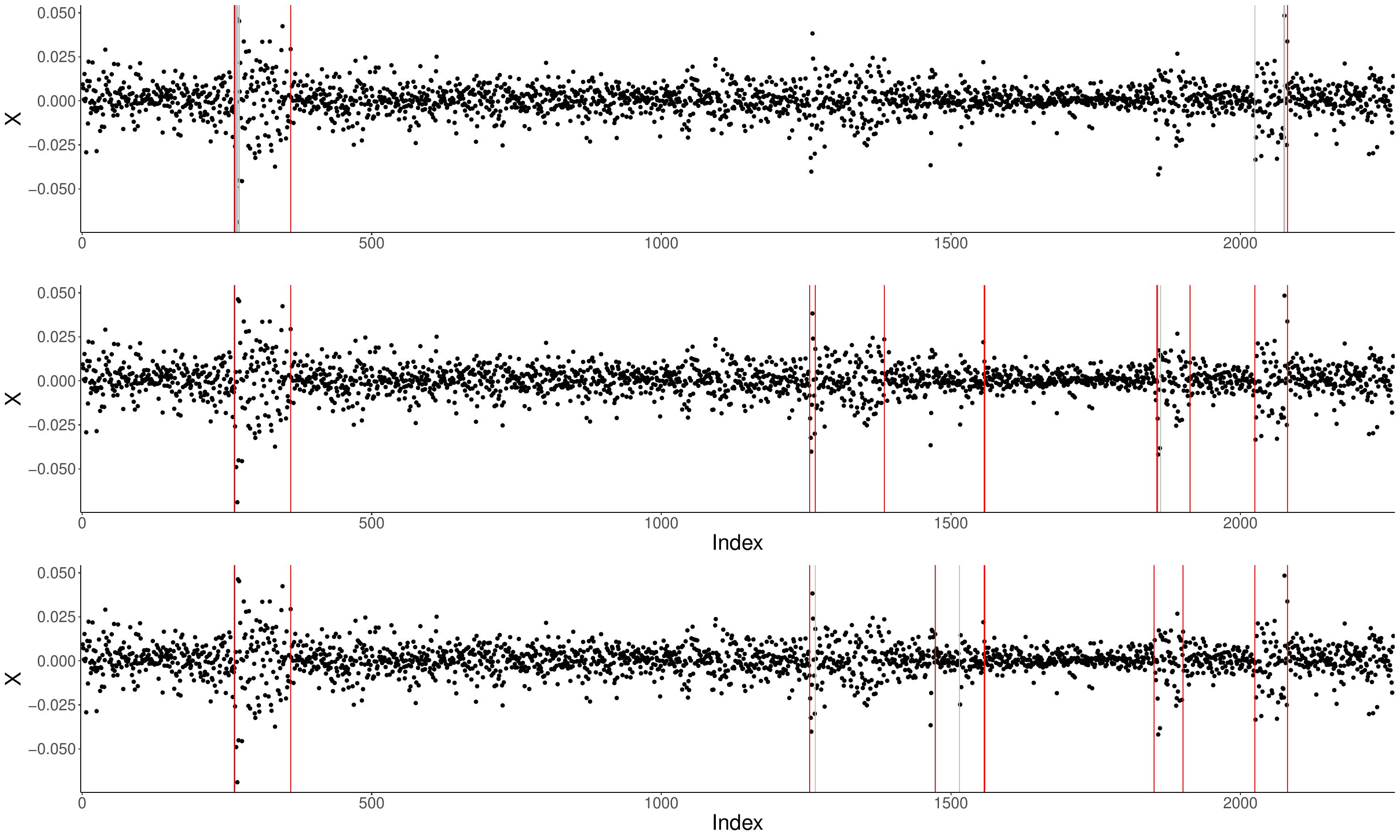}
    \caption{\small{Application of our method to financial log returns data. Changepoints are estimated using (i) CUSUM statistic with binary segmentation (top); (ii) likelihood ratio statistic with binary segmentation (middle); (iii) likelihood ratio statistic with PELT \citep{killick2012optimal}} (bottom). In each case we estimated 11 changepoints. Vertical lines indicate changepoints: those with a $p$-value smaller than 0.05 (after controlling for multiple testing using the Holm-Bonferroni method) are shown in red, and those with a $p$-value larger than 0.05 in grey.}
    \label{fig:real-data}
\end{figure}

To investigate how well our method works in a real-world setting, we applied our method to financial data (Figure \ref{fig:real-data}). The data is log returns from S\&P500 stock market data for 2264 days from July 2010 to October 2019\footnote{The data was downloaded from \url{https://www.kaggle.com/datasets/mathisjander/s-and-p500-volatility-prediction-time-series-data}}. We can see from Figure \ref{fig:real-data} that the data appears to have zero mean and several changes in variance. We estimated changepoints using three different methods: binary segmentation with the CUSUM statistic (top) and the likelihood ratio statistic (middle) and the penalized likelihood method PELT \citep{killick2012optimal} (bottom), and calculated $p$-values for each estimated changepoint. We set thresholds so that in each case we estimated 11 changepoints, and set $h = 50$. For the cases where we had to estimate $p$-values, we used a Gaussian process with $N = 100$. On each plot estimated changepoints are shown by vertical lines, with red lines corresponding to estimated changepoint locations whose $p$-values were smaller than $0.05$ (after controlling for multiple testing using the Holm-Bonferroni method) and grey lines corresponding to $p$-values which were not significant. Using the likelihood ratio statistic generally seems to give better results than CUSUM, where there are more instances of multiple changepoints being estimated close together. However, these spurious changepoints generally have non-significant $p$-values associated with them, showing that our method for post-selection inference is effective at distinguishing true from false changepoint estimates.

\section{Discussion} \label{sec:discussion}

In this paper we have proposed a test for the piecewise constant change in variance model. We have developed methods for post-selection inference, which allow us to quantify uncertainty by computing $p$-values for estimated changes, and we have shown that our methods work on simulated and real data.

We have worked on the assumption that the data is i.i.d. Gaussian except for abrupt changes in variance. In practice, less restrictive models may be more appropriate in many cases, for example allowing the mean to change as well as the variance, or allowing autocorrelation between data points. Further work in this area could include extending or adapting our approach to apply to such models.

The strategies we have developed using Gaussian processes and importance sampling may be useful for other similar situations where we conduct post-selection inference but cannot characterise the selection event explicitly. More efficient implementations may be possible, for example using Bayesian optimisation \cite[]{frazier2018tutorial} to adaptively choose $\phi$ values to consider based on some expected improvement in some appropriate measure of accuracy of the estimate of the $p$-value.

\section*{Acknowledgements}

This work was supported by EPSRC grant number EP/V053590/1.

\bibliography{references}


\begin{thebibliography}{39}
\ifx \bisbn   \undefined \def \bisbn  #1{ISBN #1}\fi
\ifx \binits  \undefined \def \binits#1{#1}\fi
\ifx \bauthor  \undefined \def \bauthor#1{#1}\fi
\ifx \batitle  \undefined \def \batitle#1{#1}\fi
\ifx \bjtitle  \undefined \def \bjtitle#1{#1}\fi
\ifx \bvolume  \undefined \def \bvolume#1{\textbf{#1}}\fi
\ifx \byear  \undefined \def \byear#1{#1}\fi
\ifx \bissue  \undefined \def \bissue#1{#1}\fi
\ifx \bfpage  \undefined \def \bfpage#1{#1}\fi
\ifx \blpage  \undefined \def \blpage #1{#1}\fi
\ifx \burl  \undefined \def \burl#1{\textsf{#1}}\fi
\ifx \doiurl  \undefined \def \doiurl#1{\url{https://doi.org/#1}}\fi
\ifx \betal  \undefined \def \betal{\textit{et al.}}\fi
\ifx \binstitute  \undefined \def \binstitute#1{#1}\fi
\ifx \binstitutionaled  \undefined \def \binstitutionaled#1{#1}\fi
\ifx \bctitle  \undefined \def \bctitle#1{#1}\fi
\ifx \beditor  \undefined \def \beditor#1{#1}\fi
\ifx \bpublisher  \undefined \def \bpublisher#1{#1}\fi
\ifx \bbtitle  \undefined \def \bbtitle#1{#1}\fi
\ifx \bedition  \undefined \def \bedition#1{#1}\fi
\ifx \bseriesno  \undefined \def \bseriesno#1{#1}\fi
\ifx \blocation  \undefined \def \blocation#1{#1}\fi
\ifx \bsertitle  \undefined \def \bsertitle#1{#1}\fi
\ifx \bsnm \undefined \def \bsnm#1{#1}\fi
\ifx \bsuffix \undefined \def \bsuffix#1{#1}\fi
\ifx \bparticle \undefined \def \bparticle#1{#1}\fi
\ifx \barticle \undefined \def \barticle#1{#1}\fi
\bibcommenthead
\ifx \bconfdate \undefined \def \bconfdate #1{#1}\fi
\ifx \botherref \undefined \def \botherref #1{#1}\fi
\ifx \url \undefined \def \url#1{\textsf{#1}}\fi
\ifx \bchapter \undefined \def \bchapter#1{#1}\fi
\ifx \bbook \undefined \def \bbook#1{#1}\fi
\ifx \bcomment \undefined \def \bcomment#1{#1}\fi
\ifx \oauthor \undefined \def \oauthor#1{#1}\fi
\ifx \citeauthoryear \undefined \def \citeauthoryear#1{#1}\fi
\ifx \endbibitem  \undefined \def \endbibitem {}\fi
\ifx \bconflocation  \undefined \def \bconflocation#1{#1}\fi
\ifx \arxivurl  \undefined \def \arxivurl#1{\textsf{#1}}\fi
\csname PreBibitemsHook\endcsname

%%% 1
\bibitem[\protect\citeauthoryear{Anastasiou et~al.}{2022}]{anastasiou2022new}
\begin{barticle}
\bauthor{\bsnm{Anastasiou}, \binits{A.}},
\bauthor{\bsnm{Cribben}, \binits{I.}},
\bauthor{\bsnm{Fryzlewicz}, \binits{P.}}:
\batitle{Cross-covariate isolate detect: A new change-point method for estimating dynamic functional connectivity}.
\bjtitle{Medical Image Analysis}
\bvolume{75},
\bfpage{102252}
(\byear{2022})
\end{barticle}
\endbibitem

%%% 2
\bibitem[\protect\citeauthoryear{Amiri et~al.}{2015}]{amiri2015probabilistic}
\begin{barticle}
\bauthor{\bsnm{Amiri}, \binits{A.}},
\bauthor{\bsnm{Niaki}, \binits{S.T.A.}},
\bauthor{\bsnm{Moghadam}, \binits{A.T.}}:
\batitle{A probabilistic artificial neural network-based procedure for variance change point estimation}.
\bjtitle{Soft Computing}
\bvolume{19},
\bfpage{691}--\blpage{700}
(\byear{2015})
\end{barticle}
\endbibitem

%%% 3
\bibitem[\protect\citeauthoryear{Bayarri and Berger}{1999}]{bayarri1999quantifying}
\begin{barticle}
\bauthor{\bsnm{Bayarri}, \binits{M.}},
\bauthor{\bsnm{Berger}, \binits{J.O.}}:
\batitle{Quantifying surprise in the data and model verification}.
\bjtitle{Bayesian Statistics}
\bvolume{6},
\bfpage{53}--\blpage{82}
(\byear{1999})
\end{barticle}
\endbibitem

%%% 4
\bibitem[\protect\citeauthoryear{Baranowski et~al.}{2019}]{baranowski2019narrowest}
\begin{barticle}
\bauthor{\bsnm{Baranowski}, \binits{R.}},
\bauthor{\bsnm{Chen}, \binits{Y.}},
\bauthor{\bsnm{Fryzlewicz}, \binits{P.}}:
\batitle{Narrowest-over-threshold detection of multiple change points and change-point-like features}.
\bjtitle{Journal of the Royal Statistical Society: Series B}
\bvolume{81}(\bissue{3}),
\bfpage{649}--\blpage{672}
(\byear{2019})
\end{barticle}
\endbibitem

%%% 5
\bibitem[\protect\citeauthoryear{Barry and Hartigan}{1993}]{barry1993bayesian}
\begin{barticle}
\bauthor{\bsnm{Barry}, \binits{D.}},
\bauthor{\bsnm{Hartigan}, \binits{J.A.}}:
\batitle{A {B}ayesian analysis for change point problems}.
\bjtitle{Journal of the American Statistical Association}
\bvolume{88}(\bissue{421}),
\bfpage{309}--\blpage{319}
(\byear{1993})
\end{barticle}
\endbibitem

%%% 6
\bibitem[\protect\citeauthoryear{Caron et~al.}{2012}]{caron2012line}
\begin{barticle}
\bauthor{\bsnm{Caron}, \binits{F.}},
\bauthor{\bsnm{Doucet}, \binits{A.}},
\bauthor{\bsnm{Gottardo}, \binits{R.}}:
\batitle{On-line changepoint detection and parameter estimation with application to genomic data}.
\bjtitle{Statistics and Computing}
\bvolume{22},
\bfpage{579}--\blpage{595}
(\byear{2012})
\end{barticle}
\endbibitem

%%% 7
\bibitem[\protect\citeauthoryear{Carrington and Fearnhead}{2025}]{carrington2023improving}
\begin{barticle}
\bauthor{\bsnm{Carrington}, \binits{R.}},
\bauthor{\bsnm{Fearnhead}, \binits{P.}}:
\batitle{Improving power by conditioning on less in post-selection inference for changepoints}.
\bjtitle{Statistics and Computing}
\bvolume{35}(\bissue{1}),
\bfpage{1}--\blpage{23}
(\byear{2025})
\end{barticle}
\endbibitem

%%% 8
\bibitem[\protect\citeauthoryear{Chen et~al.}{2023}]{chen2023quantifying}
\begin{barticle}
\bauthor{\bsnm{Chen}, \binits{Y.T.}},
\bauthor{\bsnm{Jewell}, \binits{S.W.}},
\bauthor{\bsnm{Witten}, \binits{D.M.}}:
\batitle{Quantifying uncertainty in spikes estimated from calcium imaging data}.
\bjtitle{Biostatistics}
\bvolume{24}(\bissue{2}),
\bfpage{481}--\blpage{501}
(\byear{2023})
\end{barticle}
\endbibitem

%%% 9
\bibitem[\protect\citeauthoryear{Cappello and Padilla}{2023}]{cappello2023bayesian}
\begin{botherref}
\oauthor{\bsnm{Cappello}, \binits{L.}},
\oauthor{\bsnm{Padilla}, \binits{O.H.M.}}:
Bayesian variance change point detection with credible sets.
arXiv preprint arXiv:2211.14097
(2023)
\end{botherref}
\endbibitem

%%% 10
\bibitem[\protect\citeauthoryear{Duy and Takeuchi}{2022}]{duy2022more}
\begin{barticle}
\bauthor{\bsnm{Duy}, \binits{V.N.L.}},
\bauthor{\bsnm{Takeuchi}, \binits{I.}}:
\batitle{More powerful conditional selective inference for generalized lasso by parametric programming}.
\bjtitle{The Journal of Machine Learning Research}
\bvolume{23}(\bissue{1}),
\bfpage{13544}--\blpage{13580}
(\byear{2022})
\end{barticle}
\endbibitem

%%% 11
\bibitem[\protect\citeauthoryear{Eichinger and Kirch}{2018}]{eichinger2018mosum}
\begin{barticle}
\bauthor{\bsnm{Eichinger}, \binits{B.}},
\bauthor{\bsnm{Kirch}, \binits{C.}}:
\batitle{A mosum procedure for the estimation of multiple random change points}.
\bjtitle{Bernoulli}
\bvolume{24},
\bfpage{526}--\blpage{564}
(\byear{2018})
\end{barticle}
\endbibitem

%%% 12
\bibitem[\protect\citeauthoryear{Fearnhead}{2006}]{fearnhead2006exact}
\begin{barticle}
\bauthor{\bsnm{Fearnhead}, \binits{P.}}:
\batitle{Exact and efficient {B}ayesian inference for multiple changepoint problems}.
\bjtitle{Statistics and Computing}
\bvolume{16},
\bfpage{203}--\blpage{213}
(\byear{2006})
\end{barticle}
\endbibitem

%%% 13
\bibitem[\protect\citeauthoryear{Frick et~al.}{2014}]{frick2014multiscale}
\begin{barticle}
\bauthor{\bsnm{Frick}, \binits{K.}},
\bauthor{\bsnm{Munk}, \binits{A.}},
\bauthor{\bsnm{Sieling}, \binits{H.}}:
\batitle{Multiscale change point inference}.
\bjtitle{Journal of the Royal Statistical Society: Series B}
\bvolume{76}(\bissue{3}),
\bfpage{495}--\blpage{580}
(\byear{2014})
\end{barticle}
\endbibitem

%%% 14
\bibitem[\protect\citeauthoryear{Fearnhead and Rigaill}{2020}]{fearnhead2020relating}
\begin{barticle}
\bauthor{\bsnm{Fearnhead}, \binits{P.}},
\bauthor{\bsnm{Rigaill}, \binits{G.}}:
\batitle{Relating and comparing methods for detecting changes in mean}.
\bjtitle{Stat}
\bvolume{9}(\bissue{1}),
\bfpage{291}
(\byear{2020})
\end{barticle}
\endbibitem

%%% 15
\bibitem[\protect\citeauthoryear{Frazier}{2018}]{frazier2018tutorial}
\begin{botherref}
\oauthor{\bsnm{Frazier}, \binits{P.I.}}:
A tutorial on {B}ayesian optimization.
arXiv preprint arXiv:1807.02811
(2018)
\end{botherref}
\endbibitem

%%% 16
\bibitem[\protect\citeauthoryear{Fryzlewicz}{2014}]{fryzlewicz2014wild}
\begin{barticle}
\bauthor{\bsnm{Fryzlewicz}, \binits{P.}}:
\batitle{Wild binary segmentation for multiple change-point detection}.
\bjtitle{The Annals of Statistics}
\bvolume{42}(\bissue{6}),
\bfpage{2243}--\blpage{2281}
(\byear{2014})
\end{barticle}
\endbibitem

%%% 17
\bibitem[\protect\citeauthoryear{Fryzlewicz}{2023}]{fryzlewicz2023narrowest}
\begin{botherref}
\oauthor{\bsnm{Fryzlewicz}, \binits{P.}}:
Narrowest significance pursuit: inference for multiple change-points in linear models.
Journal of the American Statistical Association,
1--14
(2023)
\end{botherref}
\endbibitem

%%% 18
\bibitem[\protect\citeauthoryear{Fryzlewicz}{2024}]{fryzlewicz2024robust}
\begin{botherref}
\oauthor{\bsnm{Fryzlewicz}, \binits{P.}}:
Robust narrowest significance pursuit: Inference for multiple change-points in the median.
Journal of Business \& Economic Statistics,
1--14
(2024)
\end{botherref}
\endbibitem

%%% 19
\bibitem[\protect\citeauthoryear{Fithian et~al.}{2014}]{fithian2014optimal}
\begin{botherref}
\oauthor{\bsnm{Fithian}, \binits{W.}},
\oauthor{\bsnm{Sun}, \binits{D.}},
\oauthor{\bsnm{Taylor}, \binits{J.}}:
Optimal inference after model selection.
arXiv preprint arXiv:1410.2597
(2014)
\end{botherref}
\endbibitem

%%% 20
\bibitem[\protect\citeauthoryear{Hyun et~al.}{2021}]{hyun2021post}
\begin{barticle}
\bauthor{\bsnm{Hyun}, \binits{S.}},
\bauthor{\bsnm{Lin}, \binits{K.Z.}},
\bauthor{\bsnm{G'Sell}, \binits{M.}},
\bauthor{\bsnm{Tibshirani}, \binits{R.J.}}:
\batitle{Post-selection inference for changepoint detection algorithms with application to copy number variation data}.
\bjtitle{Biometrics}
\bvolume{77}(\bissue{3}),
\bfpage{1037}--\blpage{1049}
(\byear{2021})
\end{barticle}
\endbibitem

%%% 21
\bibitem[\protect\citeauthoryear{Inclan and Tiao}{1994}]{inclan1994use}
\begin{barticle}
\bauthor{\bsnm{Inclan}, \binits{C.}},
\bauthor{\bsnm{Tiao}, \binits{G.C.}}:
\batitle{Use of cumulative sums of squares for retrospective detection of changes of variance}.
\bjtitle{Journal of the American Statistical Association}
\bvolume{89}(\bissue{427}),
\bfpage{913}--\blpage{923}
(\byear{1994})
\end{barticle}
\endbibitem

%%% 22
\bibitem[\protect\citeauthoryear{Jewell et~al.}{2022}]{jewell2022testing}
\begin{barticle}
\bauthor{\bsnm{Jewell}, \binits{S.}},
\bauthor{\bsnm{Fearnhead}, \binits{P.}},
\bauthor{\bsnm{Witten}, \binits{D.}}:
\batitle{Testing for a change in mean after changepoint detection}.
\bjtitle{Journal of the Royal Statistical Society: Series B}
\bvolume{84}(\bissue{4}),
\bfpage{1082}--\blpage{1104}
(\byear{2022})
\end{barticle}
\endbibitem

%%% 23
\bibitem[\protect\citeauthoryear{Jiang et~al.}{2023}]{jiang2023time}
\begin{barticle}
\bauthor{\bsnm{Jiang}, \binits{F.}},
\bauthor{\bsnm{Zhao}, \binits{Z.}},
\bauthor{\bsnm{Shao}, \binits{X.}}:
\batitle{Time series analysis of covid-19 infection curve: A change-point perspective}.
\bjtitle{Journal of econometrics}
\bvolume{232}(\bissue{1}),
\bfpage{1}--\blpage{17}
(\byear{2023})
\end{barticle}
\endbibitem

%%% 24
\bibitem[\protect\citeauthoryear{Kov{\'a}cs et~al.}{2023}]{kovacs2023seeded}
\begin{barticle}
\bauthor{\bsnm{Kov{\'a}cs}, \binits{S.}},
\bauthor{\bsnm{B{\"u}hlmann}, \binits{P.}},
\bauthor{\bsnm{Li}, \binits{H.}},
\bauthor{\bsnm{Munk}, \binits{A.}}:
\batitle{Seeded binary segmentation: a general methodology for fast and optimal changepoint detection}.
\bjtitle{Biometrika}
\bvolume{110}(\bissue{1}),
\bfpage{249}--\blpage{256}
(\byear{2023})
\end{barticle}
\endbibitem

%%% 25
\bibitem[\protect\citeauthoryear{Killick et~al.}{2012}]{killick2012optimal}
\begin{barticle}
\bauthor{\bsnm{Killick}, \binits{R.}},
\bauthor{\bsnm{Fearnhead}, \binits{P.}},
\bauthor{\bsnm{Eckley}, \binits{I.A.}}:
\batitle{Optimal detection of changepoints with a linear computational cost}.
\bjtitle{Journal of the American Statistical Association}
\bvolume{107}(\bissue{500}),
\bfpage{1590}--\blpage{1598}
(\byear{2012})
\end{barticle}
\endbibitem

%%% 26
\bibitem[\protect\citeauthoryear{Kuchibhotla et~al.}{2022}]{kuchibhotla2022post}
\begin{barticle}
\bauthor{\bsnm{Kuchibhotla}, \binits{A.K.}},
\bauthor{\bsnm{Kolassa}, \binits{J.E.}},
\bauthor{\bsnm{Kuffner}, \binits{T.A.}}:
\batitle{Post-selection inference}.
\bjtitle{Annual Review of Statistics and Its Application}
\bvolume{9},
\bfpage{505}--\blpage{527}
(\byear{2022})
\end{barticle}
\endbibitem

%%% 27
\bibitem[\protect\citeauthoryear{Li et~al.}{2016}]{li2016fdr}
\begin{barticle}
\bauthor{\bsnm{Li}, \binits{H.}},
\bauthor{\bsnm{Munk}, \binits{A.}},
\bauthor{\bsnm{Sieling}, \binits{H.}}:
\batitle{Fdr-control in multiscale change-point segmentation}.
\bjtitle{Electronic Journal of Statistics}
\bvolume{10},
\bfpage{918}--\blpage{959}
(\byear{2016})
\end{barticle}
\endbibitem

%%% 28
\bibitem[\protect\citeauthoryear{Muggeo and Adelfio}{2011}]{muggeo2011efficient}
\begin{barticle}
\bauthor{\bsnm{Muggeo}, \binits{V.M.}},
\bauthor{\bsnm{Adelfio}, \binits{G.}}:
\batitle{Efficient change point detection for genomic sequences of continuous measurements}.
\bjtitle{Bioinformatics}
\bvolume{27}(\bissue{2}),
\bfpage{161}--\blpage{166}
(\byear{2011})
\end{barticle}
\endbibitem

%%% 29
\bibitem[\protect\citeauthoryear{Mehrizi and Chenouri}{2021}]{mehrizi2021valid}
\begin{botherref}
\oauthor{\bsnm{Mehrizi}, \binits{R.V.}},
\oauthor{\bsnm{Chenouri}, \binits{S.}}:
Valid post-detection inference for change points identified using trend filtering.
arXiv preprint arXiv:2104.12022
(2021)
\end{botherref}
\endbibitem

%%% 30
\bibitem[\protect\citeauthoryear{Maidstone et~al.}{2017}]{maidstone2017optimal}
\begin{barticle}
\bauthor{\bsnm{Maidstone}, \binits{R.}},
\bauthor{\bsnm{Hocking}, \binits{T.}},
\bauthor{\bsnm{Rigaill}, \binits{G.}},
\bauthor{\bsnm{Fearnhead}, \binits{P.}}:
\batitle{On optimal multiple changepoint algorithms for large data}.
\bjtitle{Statistics and Computing}
\bvolume{27},
\bfpage{519}--\blpage{533}
(\byear{2017})
\end{barticle}
\endbibitem

%%% 31
\bibitem[\protect\citeauthoryear{Meier et~al.}{2021}]{meier2021mosum}
\begin{barticle}
\bauthor{\bsnm{Meier}, \binits{A.}},
\bauthor{\bsnm{Kirch}, \binits{C.}},
\bauthor{\bsnm{Cho}, \binits{H.}}:
\batitle{mosum: A package for moving sums in change point analysis}.
\bjtitle{Journal of Statistical Software}
\bvolume{97}(\bissue{8}),
\bfpage{1}--\blpage{42}
(\byear{2021})
\end{barticle}
\endbibitem

%%% 32
\bibitem[\protect\citeauthoryear{Nam et~al.}{2012}]{nam2012quantifying}
\begin{barticle}
\bauthor{\bsnm{Nam}, \binits{C.F.}},
\bauthor{\bsnm{Aston}, \binits{J.A.}},
\bauthor{\bsnm{Johansen}, \binits{A.M.}}:
\batitle{Quantifying the uncertainty in change points}.
\bjtitle{Journal of Time Series Analysis}
\bvolume{33}(\bissue{5}),
\bfpage{807}--\blpage{823}
(\byear{2012})
\end{barticle}
\endbibitem

%%% 33
\bibitem[\protect\citeauthoryear{Reeves et~al.}{2007}]{reeves2007review}
\begin{barticle}
\bauthor{\bsnm{Reeves}, \binits{J.}},
\bauthor{\bsnm{Chen}, \binits{J.}},
\bauthor{\bsnm{Wang}, \binits{X.L.}},
\bauthor{\bsnm{Lund}, \binits{R.}},
\bauthor{\bsnm{Lu}, \binits{Q.Q.}}:
\batitle{A review and comparison of changepoint detection techniques for climate data}.
\bjtitle{Journal of Applied Meteorology and Climatology}
\bvolume{46}(\bissue{6}),
\bfpage{900}--\blpage{915}
(\byear{2007})
\end{barticle}
\endbibitem

%%% 34
\bibitem[\protect\citeauthoryear{Shi et~al.}{2022}]{shi2022changepoint}
\begin{barticle}
\bauthor{\bsnm{Shi}, \binits{X.}},
\bauthor{\bsnm{Beaulieu}, \binits{C.}},
\bauthor{\bsnm{Killick}, \binits{R.}},
\bauthor{\bsnm{Lund}, \binits{R.}}:
\batitle{Changepoint detection: An analysis of the central {E}ngland temperature series}.
\bjtitle{Journal of Climate}
\bvolume{35}(\bissue{19}),
\bfpage{2729}--\blpage{2742}
(\byear{2022})
\end{barticle}
\endbibitem

%%% 35
\bibitem[\protect\citeauthoryear{Schr{\"o}der and Fryzlewicz}{2013}]{schroder2013adaptive}
\begin{barticle}
\bauthor{\bsnm{Schr{\"o}der}, \binits{A.L.}},
\bauthor{\bsnm{Fryzlewicz}, \binits{P.}}:
\batitle{Adaptive trend estimation in financial time series via multiscale change-point-induced basis recovery}.
\bjtitle{Statistics and Its Interface}
\bvolume{6},
\bfpage{449}--\blpage{461}
(\byear{2013})
\end{barticle}
\endbibitem

%%% 36
\bibitem[\protect\citeauthoryear{Shi et~al.}{2022}]{shi2022comparison}
\begin{barticle}
\bauthor{\bsnm{Shi}, \binits{X.}},
\bauthor{\bsnm{Gallagher}, \binits{C.}},
\bauthor{\bsnm{Lund}, \binits{R.}},
\bauthor{\bsnm{Killick}, \binits{R.}}:
\batitle{A comparison of single and multiple changepoint techniques for time series data}.
\bjtitle{Computational Statistics \& Data Analysis}
\bvolume{170},
\bfpage{107433}
(\byear{2022})
\end{barticle}
\endbibitem

%%% 37
\bibitem[\protect\citeauthoryear{Scott and Knott}{1974}]{scott1974scott}
\begin{barticle}
\bauthor{\bsnm{Scott}, \binits{A.J.}},
\bauthor{\bsnm{Knott}, \binits{M.}}:
\batitle{A cluster analysis method for grouping means in the analysis of variance}.
\bjtitle{Biometrics}
\bvolume{30},
\bfpage{507}--\blpage{512}
(\byear{1974})
\end{barticle}
\endbibitem

%%% 38
\bibitem[\protect\citeauthoryear{Saha et~al.}{2024}]{saha2022inferring}
\begin{botherref}
\oauthor{\bsnm{Saha}, \binits{A.}},
\oauthor{\bsnm{Witten}, \binits{D.}},
\oauthor{\bsnm{Bien}, \binits{J.}}:
Inferring independent sets of {G}aussian variables after thresholding correlations.
Journal of the American Statistical Association,
1--12
(2024)
\end{botherref}
\endbibitem

%%% 39
\bibitem[\protect\citeauthoryear{Williams and Rasmussen}{2005}]{williams2006gaussian}
\begin{bbook}
\bauthor{\bsnm{Williams}, \binits{C.K.}},
\bauthor{\bsnm{Rasmussen}, \binits{C.E.}}:
\bbtitle{Gaussian Processes for Machine Learning}.
\bpublisher{MIT press},
\blocation{Cambridge, MA}
(\byear{2005})
\end{bbook}
\endbibitem

\end{thebibliography}
\newpage
\setcounter{page}{1}

\begin{center}
\Large
Supplementary Material for ``Post-selection inference for quantifying uncertainty in changes in variance'' \\ 
Rachel Carrington and Paul Fearnhead
\end{center}

\appendix

\section{Details for Section \ref{sec:lr-psi}}

We can show that
\begin{equation*}
    \Lambda_{s,e}(\tau, \phi) = \log \left( \frac{(A_{s, e} \phi + B_{s,e})^{s - e + 1}}{(A_{s,\tau} \phi + B_{s,\tau})^{\tau - s + 1} ((A_{s,e} - A_{s,\tau})\phi + B_{s,e} - B_{s,\tau})^{e - \tau}} \right),
\end{equation*}
where $A_{s,e}$ and $B_{s,e}$ are constants depending on $s$, $e$, $h$, and $\phi_{obs}$; specifically

\begin{equation*}
    A_{s, e} = \begin{cases}
       0 & s, e \leq \hat{\tau} - h \\
       \frac{1}{\phi_{obs}} \sum_{t=\hat{\tau} - h + 1}^e X_t^2 & s \leq \hat{\tau} - h < e \leq \hat{\tau} \\
       \frac{1}{\phi_{obs}} \sum_{t = \hat{\tau} - h + 1}^{\hat{\tau}} X_t^2 - \frac{1}{1 - \phi_{obs}} \sum_{t = \hat{\tau} + 1}^e X_t^2 & s \leq \hat{\tau} - h, \hat{\tau} < e < \hat{\tau} + h \\
       \frac{1}{\phi_{obs}} \sum_{t=\hat{\tau} - h + 1}^{\hat{\tau}} X_t^2 - \frac{1}{1 - \phi_{obs}} \sum_{t=\hat{\tau} + 1}^{\hat{\tau} + h} X_t^2 & s \leq \hat{\tau} - h, \hat{\tau} + h < e \\
       \frac{1}{\phi_{obs}} \sum_{t=s}^e X_t^2 & \hat{\tau} - h < s < e \leq \hat{\tau} \\
       \frac{1}{\phi_{obs}} \sum_{t=s}^{\hat{\tau}} X_t^2 - \frac{1}{1 - \phi_{obs}} \sum_{t=\hat{\tau} + 1}^{\min{\{\hat{\tau} + h, e\}}} X_t^2 & \hat{\tau} - h < s \leq \hat{\tau} < e \\
       - \frac{1}{1 - \phi_{obs}} \sum_{t=s}^{\min{\{\hat{\tau} + h, e\}}} X_t^2 & \hat{\tau} < s \leq \hat{\tau} + h, e > s \\
       0 & s, e > \hat{\tau} + h
    \end{cases}
\end{equation*}

\begin{equation*}
    B_{s, e} = \begin{cases}
        \sum_{t=s}^{\min{\{e, \hat{\tau} - h\}}} X_t^2 & s \leq \hat{\tau} - h, e \leq \hat{\tau} \\
        \sum_{t=s}^{\hat{\tau} - h} X_t^2 + \frac{1}{1 - \phi_{obs}} \sum_{t=\hat{\tau} + 1}^e X_t^2 & s \leq \hat{\tau} - h < \hat{\tau} < e \leq \hat{\tau} + h \\
        \sum_{t=s}^{\hat{\tau} - h} X_t^2 + \frac{1}{1 - \phi_{obs}} \sum_{t=\hat{\tau} + 1}^{\hat{\tau} + h} X_t^2 + \sum_{t=\hat{\tau} + h + 1}^e X_t^2 & s \leq \hat{\tau} - h < \hat{\tau} + h < e \\
        0 & \hat{\tau} - h < s < e \leq \hat{\tau} \\
        \frac{1}{1 - \phi_{obs}} \sum_{t=\hat{\tau} + 1}^e X_t^2 & \hat{\tau} - h < s \leq \hat{\tau} < e \leq \hat{\tau} + h \\
        \frac{1}{1 - \phi_{obs}} \sum_{t=\hat{\tau} + 1}^{\hat{\tau} + h} X_t^2 + \sum_{t=\hat{\tau} + h + 1}^e X_t^2 & \hat{\tau} - h < s \leq \hat{\tau} < \hat{\tau} + h < e \\
        \frac{1}{1 - \phi_{obs}} \sum_{t=s}^e X_t^2 & \hat{\tau} < s < e \leq \hat{\tau} + h \\
        \frac{1}{1 - \phi_{obs}} \sum_{t=s}^{\hat{\tau} + h} X_t^2 + \sum_{t=\hat{\tau} + h + 1}^e X_t^2 & \hat{\tau} < s \leq \hat{\tau} + h < e \\
        \sum_{t=s}^e X_t^2 & s, e > \hat{\tau} + h
        \end{cases}
\end{equation*}

\section{Proof of Lemma \ref{lem:independence}}

We want to show that $W_1^l, \ldots, W_{h-1}^l, W_1^r, \ldots, W_{h-1}^r, C_0^2, \phi$ are independent. To do this we calculate the joint density of these variables and show that it can be factorized into functions of each variable.

Let $g_i$ ($i = 1, \ldots, 2h)$ denote the function such that $X_{\hat{\tau} - h + i}^2 = g_i(\boldsymbol{W}, \phi, C_0^2)$, and let $\boldsymbol{J}$ denote the Jacobian matrix of partial derivatives
\[
    \boldsymbol{J} = \left( \begin{array}{cccccccc}
        \frac{\partial X_{\hat{\tau} - h + 1}^2}{\partial W_1^l} & \ldots & \frac{\partial X_{\hat{\tau} - h + 1}^2}{\partial W_{h-1}^l} & \frac{\partial X_{\hat{\tau} - h + 1}^2}{\partial W_1^r} & \ldots & \frac{\partial{X}_{\hat{\tau} - h + 1}^2}{\partial W_{h-1}^r} & \frac{\partial X_{\hat{\tau} - h + 1}^2}{\partial \phi} & \frac{\partial X_{\hat{\tau} - h + 1}^2}{\partial C_0^2} \\
        \vdots & & & & & & & \vdots \\
        \frac{\partial X_{\hat{\tau} + h}^2}{\partial W_1^l} & \ldots & \frac{\partial X_{\hat{\tau} + h}^2}{\partial W_{h-1}^l} & \frac{\partial X_{\hat{\tau} + h}^2}{\partial W_1^r} & \ldots & \frac{\partial X_{\hat{\tau} + h}^2}{\partial W_{h-1}^r} & \frac{\partial X_{\hat{\tau} + h}^2}{\partial \phi} & \frac{\partial X_{\hat{\tau} + h}^2}{\partial C_0^2}
    \end{array} \right).
\]
Then the density function of $(\boldsymbol{W}, \phi, C_0^2)$ is given by
\begin{equation*}
    f(\boldsymbol{W}, \phi, C_0^2) = \prod_{i=1}^{2h} f_{X_{\hat{\tau} - h + i}^2} (g_i^{-1}(\boldsymbol{W}, \phi, C_0^2)) |\boldsymbol{J}|.
\end{equation*}

By \eqref{eq:W} we have
\begin{equation*}
    \begin{split}
        X_{\hat{\tau} - h + i}^2 & = \begin{cases}
            (1 - W_{i-1}^l) \prod_{k=i}^{h-1} W_k^l C_0^2 \phi & i = 1, \ldots, h - 1 \\
            (1 - W_{i-1}^l) C_0^2 \phi & i = h
        \end{cases}  \\
        X_{\hat{\tau} + i}^2 & = \begin{cases}
            (1 - W_{i-1}^r) \prod_{k=i}^{h-1} W_k^r C_0^2 (1 - \phi) & i = 1, \ldots, h - 1 \\
            (1 - W_{i-1}^r) C_0^2 (1 - \phi) & i = h.
        \end{cases}
    \end{split}
\end{equation*}
Each $X_t^2 \sim Gamma \left( \frac{1}{2}, 2 \sigma^2 \right)$, so $f_{X_t}(x) = \frac{(2 \sigma^2)^{1/2}}{\Gamma(\frac{1}{2})} x^{-1/2} e^{-2 \sigma^2 x}$.

We get
\begin{equation*}
    \begin{split}
        \prod_{i=1}^{2h} f_X(g_i^{-1}(\boldsymbol{W}, \phi, C_0^2)) & = \frac{\left( 2 \sigma^2 \right)^h}{\Gamma \left( \frac{1}{2} \right)^{2h}} \left( \prod_{i=1}^{2h} X_{\hat{\tau} - h + i}^2 \right)^{-1/2} e^{-2 \sigma^2 \sum_{i=1}^{2h} X_{\hat{\tau} - h + i}^2} \\
            & = \frac{\left( 2 \sigma^2 \right)^h}{\Gamma \left( \frac{1}{2} \right)^{2h}} \prod_{i=1}^{h-1} \left(1 - W_i^l\right)^{-1/2} \prod_{i=1}^{h-1} \left(W_i^l\right)^{-i/2} \prod_{i=1}^{h-1} \left(1 - W_i^r\right)^{-1/2} \prod_{i=1}^{h-1} \left(W_i^r\right)^{-i/2} e^{-2\sigma^2 C_0^2}
    \end{split}
\end{equation*}

To calculate the determinant of $\boldsymbol{J}$, note that using elementary row operations, we can show that $|\boldsymbol{J}|$ is equal to the determinant of the matrix
\[
\begin{array}{cccccccc}
    \frac{\partial X_{\hat{\tau} - h + 1}^2}{\partial W_1^l} & \ldots & \frac{\partial X_{\hat{\tau} - h + 1}^2}{\partial W_{h-1}^l} & \frac{\partial X_{\hat{\tau} - h + 1}^2}{\partial W_1^r} & \ldots & \frac{\partial X_{\hat{\tau} - h + 1}^2}{\partial W_{h-1}^r} & \frac{\partial X_{\hat{\tau} - h + 1}^2}{\partial \phi} & \frac{\partial X_{\hat{\tau} - h + 1}^2}{\partial C_0^2} \\
    \sum_{i=1}^2 \frac{\partial X_{\hat{\tau} - h + i}^2}{\partial W_1^l} & \ldots & \sum_{i=1}^2 \frac{\partial X_{\hat{\tau} - h + i}^2}{\partial W_{h-1}^l} & \sum_{i=1}^2 \frac{\partial X_{\hat{\tau} - h + i}^2}{\partial W_1^r} & \ldots & \sum_{i=1}^2 \frac{\partial X_{\hat{\tau} - h + i}^2}{\partial W_{h-1}^r} & \sum_{i=1}^2 \frac{\partial X_{\hat{\tau} - h + i}^2}{\partial \phi} & \sum_{i=1}^2 \frac{\partial X_{\hat{\tau} - h + i}^2}{\partial C_0^2} \\
    \vdots & & \vdots & \vdots & & \vdots & \vdots & \vdots \\
    \sum_{i=1}^{h-1} \frac{\partial X_{\hat{\tau} - h + i}^2}{\partial W_1^l} & \ldots & \sum_{i=1}^{h-1} \frac{\partial X_{\hat{\tau} - h + i}^2}{\partial W_{h-1}^l} & \sum_{i=1}^{h-1} \frac{\partial X_{\hat{\tau} - h + i}^2}{\partial W_1^r} & \ldots & \sum_{i=1}^{h-1} \frac{\partial X_{\hat{\tau} - h + i}^2}{\partial W_{h-1}^r} & \sum_{i=1}^{h-1} \frac{\partial X_{\hat{\tau} - h + i}^2}{\partial \phi} & \sum_{i=1}^{h-1} \frac{\partial X_{\hat{\tau} - h + i}^2}{\partial C_0^2} \\    
    \frac{\partial X_{\hat{\tau} + 1}^2}{\partial W_1^l} & \ldots & \frac{\partial X_{\hat{\tau} + 1}^2}{\partial W_{h-1}^l} & \frac{\partial X_{\hat{\tau} + 1}^2}{\partial W_1^r} & \ldots & \frac{\partial X_{\hat{\tau} + 1}^2}{\partial W_{h-1}^r} & \frac{\partial X_{\hat{\tau} + 1}^2}{\partial \phi} & \frac{\partial X_{\hat{\tau} + 1}^2}{\partial C_0^2}  \\
    \sum_{i=1}^2 \frac{\partial X_{\hat{\tau} + i}^2}{\partial W_1^l} & \ldots & \sum_{i=1}^2 \frac{\partial X_{\hat{\tau} + i}^2}{\partial W_{h-1}^l} & \sum_{i=1}^2 \frac{\partial X_{\hat{\tau} + i}^2}{\partial W_1^r} & \ldots & \sum_{i=1}^2 \frac{\partial X_{\hat{\tau} + i}^2}{\partial W_{h-1}^r} & \sum_{i=1}^2 \frac{\partial X_{\hat{\tau} + i}^2}{\partial \phi} & \sum_{i=1}^2 \frac{\partial X_{\hat{\tau} + i}^2}{\partial C_0^2} \\
    \vdots & & \vdots & \vdots & & \vdots & \vdots & \vdots \\
    \sum_{i=1}^{h-1} \frac{\partial X_{\hat{\tau} + i}^2}{\partial W_1^l} & \ldots & \sum_{i=1}^{h-1} \frac{\partial X_{\hat{\tau} + i}^2}{\partial W_{h-1}^l} & \sum_{i=1}^{h-1} \frac{\partial X_{\hat{\tau} + i}^2}{\partial W_1^r} & \ldots & \sum_{i=1}^{h-1} \frac{\partial X_{\hat{\tau} + i}^2}{\partial W_{h-1}^r} & \sum_{i=1}^{h-1} \frac{\partial X_{\hat{\tau} + i}^2}{\partial \phi} & \sum_{i=1}^{h-1} \frac{\partial X_{\hat{\tau} + i}^2}{\partial C_0^2} \\
    \sum_{i=1}^h \frac{\partial X_{\hat{\tau} - h + i}^2}{\partial W_1^l} & \ldots & \sum_{i=1}^h \frac{\partial X_{\hat{\tau} - h + i}^2}{\partial W_{h-1}^l} & \sum_{i=1}^h \frac{\partial X_{\hat{\tau} - h + i}^2}{\partial W_1^r} & \ldots & \sum_{i=1}^h \frac{\partial X_{\hat{\tau} - h + i}^2}{\partial W_{h-1}^r} & \sum_{i=1}^h \frac{\partial X_{\hat{\tau} - h + i}^2}{\partial \phi} & \sum_{i=1}^h \frac{\partial X_{\hat{\tau} - h + i}^2}{\partial C_0^2} \\
    \sum_{i=1}^{2h} \frac{\partial X_{\hat{\tau} - h + i}^2}{\partial W_1^l} & \ldots & \sum_{i=1}^{2h} \frac{\partial X_{\hat{\tau} - h + i}^2}{\partial W_{h-1}^l} & \sum_{i=1}^{2h} \frac{\partial X_{\hat{\tau} - h + i}^2}{\partial W_1^r} & \ldots & \sum_{i=1}^{2h} \frac{\partial X_{\hat{\tau} -h + i}^2}{\partial W_{h-1}^r} & \sum_{i=1}^{2h} \frac{\partial X_{\hat{\tau} - h + i}^2}{\partial \phi} & \sum_{i=1}^{2h} \frac{\partial X_{\hat{\tau} - h + i}^2}{\partial C_0^2} \\
\end{array}.
\]

We can show that this matrix is upper triangular, which allows us to calculate the determinant as the product of its diagonal elements.
Since
\begin{equation*}
    \sum_{i=1}^k X_{\hat{\tau} - h + i}^2 = \prod_{j=k}^{h-1} W_j^l C_0^2 \phi,
\end{equation*}
it follows that $\sum_{i=1}^k \frac{\partial X_{\hat{\tau} - h + i}^2}{\partial W_j^l} = 0$ for $j < k$, and clearly we also have $\sum_{i=1}^k \frac{\partial X_{\hat{\tau} - h + i}^2}{\partial W_j^r} = 0$ for all $j$. Similarly, we can show that $\sum_{i=1}^k \frac{\partial X_{\hat{\tau} + i}^2}{\partial W_j^r} = 0$ for $j < k$, and $\sum_{i=1}^k \frac{\partial X_{\hat{\tau} + i}^2}{\partial W_j^l} = 0$ for all $k$. In the bottom row, as $\sum_{i=1}^{2h} X_{\hat{\tau} - h + i}^2 = C_0^2$, all the matrix entries except the last are 0.

We get
\begin{equation*}
  \begin{split}
    \sum_{i=1}^j \frac{\partial X_{\hat{\tau} - h + i}^2}{\partial W_j^l} & = \phi C_0^2 \prod_{k = j + 1}^{h-1} W_k^l ~~~ j, = 1, \ldots, h - 1 \\
    \sum_{i=1}^j \frac{\partial X_{\hat{\tau} + i}^2}{\partial W_j^r} & = (1 - \phi) C_0^2 \prod_{k=j+1}^{h-1} W_k^r, ~~~ j = 1, \ldots, h - 1 \\
    \sum_{i=1}^h \frac{\partial X_{\hat{\tau} - h + i}^2}{\partial \phi} & = C_0^2 \\
    \sum_{i=1}^{2h} \frac{\partial X_{\hat{\tau} - h + i}^2}{\partial C_0^2} & = 1.
  \end{split}
\end{equation*}
Hence,
\begin{equation*}
    |\boldsymbol{J}| = \prod_{i=1}^{h-1} \left( W_i^l \right)^{i - 1} \prod_{i=1}^{h-1} \left( W_i^r \right)^{i - 1} \phi^{h-1} (1 - \phi)^{h-1} \left( C_0^2 \right)^{2h - 1}.
\end{equation*}

Hence, we get
\begin{equation*}
    \begin{split}
        f(\boldsymbol{W}, \phi, C_0^2) & = \frac{(2\sigma^2)^h}{\Gamma(\frac{1}{2})^{2h}} \prod_{i=1}^{h-1} \left((W_i^l)^i(1 - W_i^l)\right)^{-1/2} \prod_{i=1}^{h-1} \left((W_i^r)^i)(1 - W_i^r)\right)^{-1/2} \left(\phi (1 - \phi)\right)^{-h/2} C_0^{-2h} e^{-2 \sigma^2 C_0^2}.
    \end{split}
\end{equation*}
This can be factorized as $h_0 \prod_{i=1}^{h-1} h_i(W_i^l) \prod_{i=1}^{h-1} h_i(W_i^r) h_{\phi}(\phi) h_C(C_0^2)$. It is also verifiable that all these variables have domain $[0, 1]$ (except for $C_0^2$ which is $[0, \infty)$) regardless of the values of the others. Hence these variables are all independent.

\section{Additional simulations} \label{sec:add-sims}

Here we include some additional simulations to those in Section \ref{sec:sims}, where we include different changepoint algorithms and parameter values.

\subsection{Detecting changes using CUSUM} \label{sec:add-cusum}

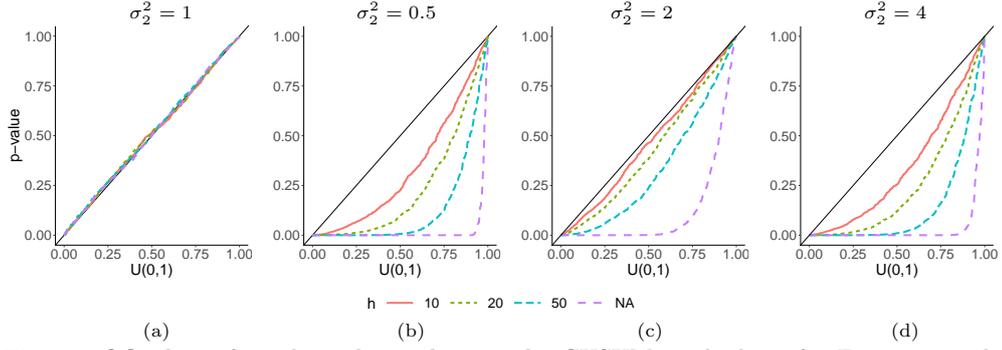
\begin{figure}
    \centering
    \includegraphics[width=\linewidth]{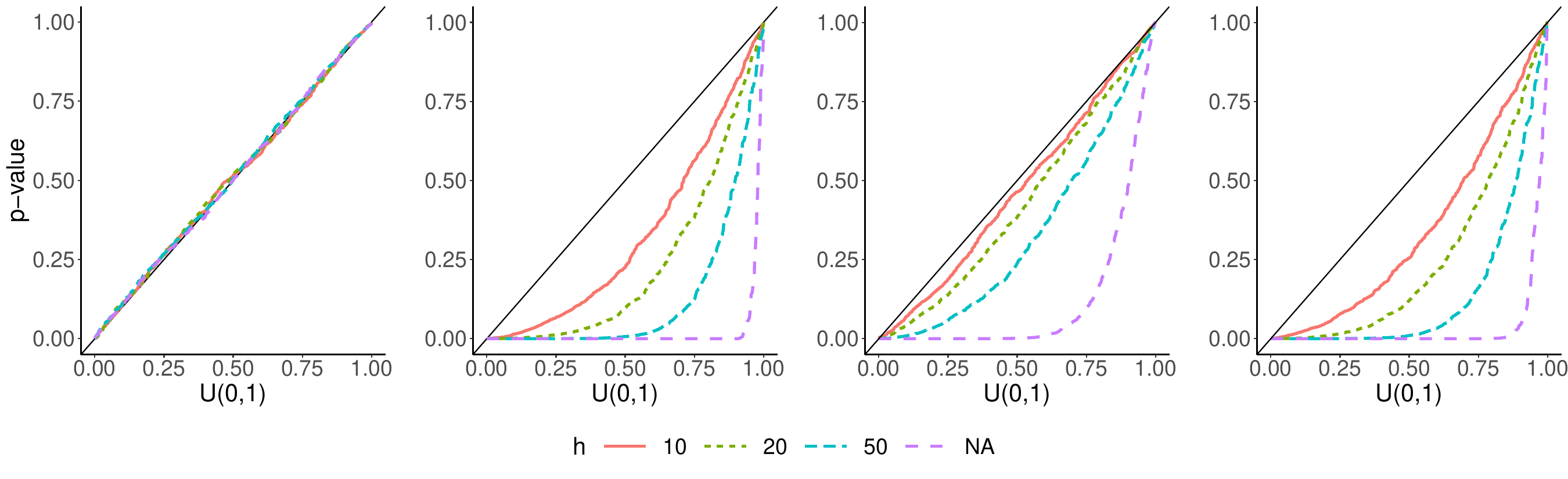}
        \put(-325, 115){\footnotesize{$\sigma_2^2 = 1$}}
            \put(-320, -5){\footnotesize{(a)}}
        \put(-240, 115){\footnotesize{$\sigma_2^2 = 0.25$}}
            \put(-225, -5){\footnotesize{(b)}}
        \put(-145, 115){\footnotesize{$\sigma_2^2 = 2$}}
            \put(-135, -5){\footnotesize{(c)}}
        \put(-50, 115){\footnotesize{$\sigma_2^2 = 4$}}
            \put(-40, -5){\footnotesize{(d)}}
    \caption{\small{QQ plots of $p$-values obtained using the CUSUM method, as for Figure \ref{fig:cusum}, with $T = 1000$. Data is simulated from a model with a single change at $\tau = T/2$, with a pre-change variance of 1, and post-change variance $\sigma_2^2$.}}
    \label{fig:cusum-1000}
\end{figure}

\begin{figure}
    \centering
    \includegraphics[width=\linewidth]{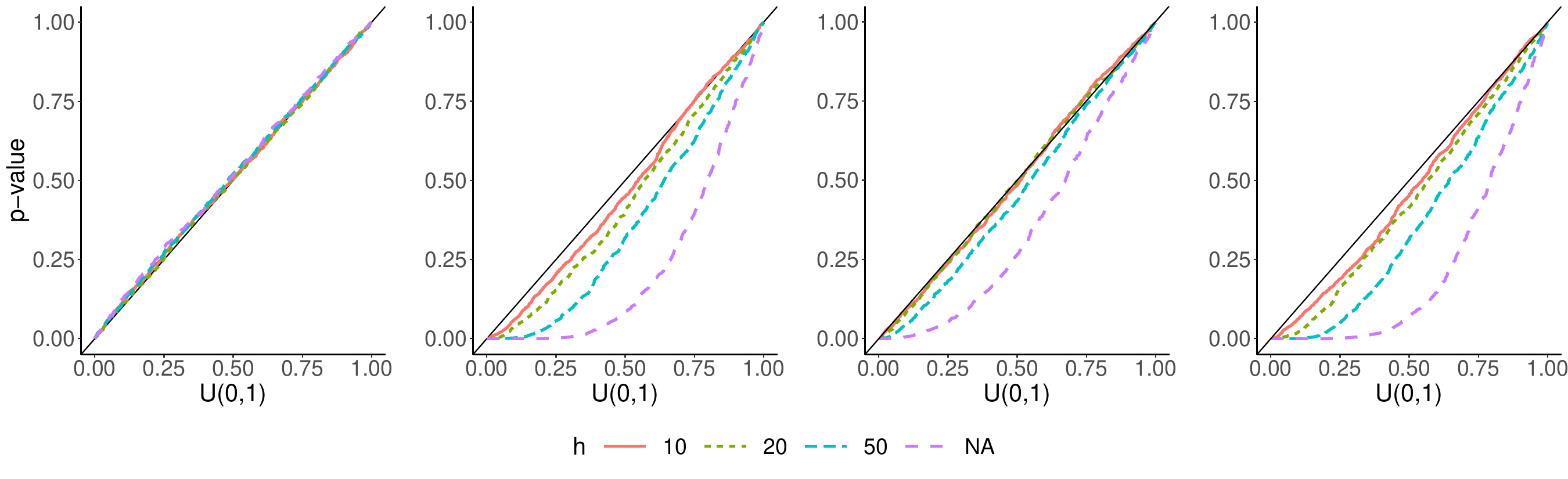}
        \put(-325, 115){\footnotesize{$\sigma_2^2 = 1$}}
            \put(-320, -5){\footnotesize{(a)}}
        \put(-240, 115){\footnotesize{$\sigma_2^2 = 0.5$}}
            \put(-225, -5){\footnotesize{(b)}}
        \put(-145, 115){\footnotesize{$\sigma_2^2 = 2$}}
            \put(-135, -5){\footnotesize{(c)}}
        \put(-50, 115){\footnotesize{$\sigma_2^2 = 4$}}
            \put(-40, -5){\footnotesize{(d)}}
    \caption{\small{QQ plots of $p$-values obtained using the CUSUM method, as for Figure \ref{fig:cusum}. Data is simulated from a model with a single change at $\tau = T/2$, with a pre-change variance of 1, and post-change variance $\sigma_2^2$. Here, we use wild binary segmentation to estimate changepoints.}}
    \label{fig:cusum-wbs-200}
\end{figure}
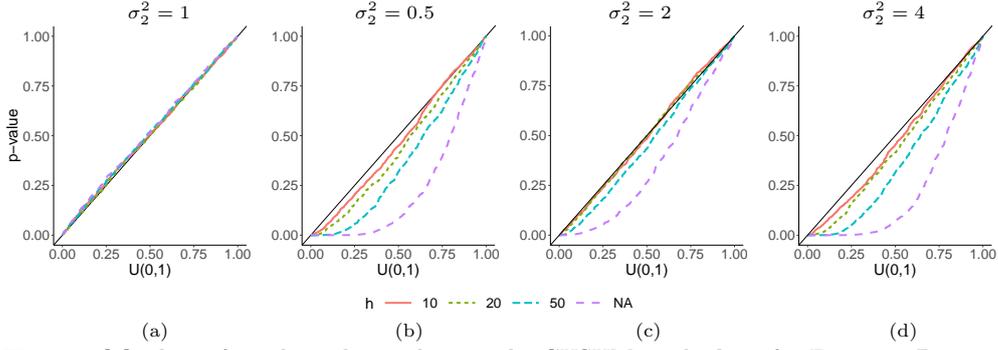

Figure \ref{fig:cusum} showed QQ plots of $p$-values obtained for simulated data sets, where changepoints were estimated using binary segmentation with the CUSUM statistic. Here, we present similar results for two different scenarios: in Figure \ref{fig:cusum-1000} we simulate data with $T = 1000$ data points (compared to $T = 200$ in Figure \ref{fig:cusum}) and once again estimate changepoints using binary segmentation; in Figure \ref{fig:cusum-wbs-200} we set $T = 200$ and estimate changepoints using wild binary segmentation.

In each figure, panel (a) shows the results when data was simulated under $H_0$; panels (b)--(d) show $p$-values obtained when we simulated from a model with a single changepoint at $\tau = T/2$, with three different values of the post-change variance. In each case we see a similar pattern of results to Figure \ref{fig:cusum}.

\subsection{Sampling with and without stratification} \label{sec:strat}

\begin{figure}
    \centering
    \includegraphics[width=0.7\linewidth]{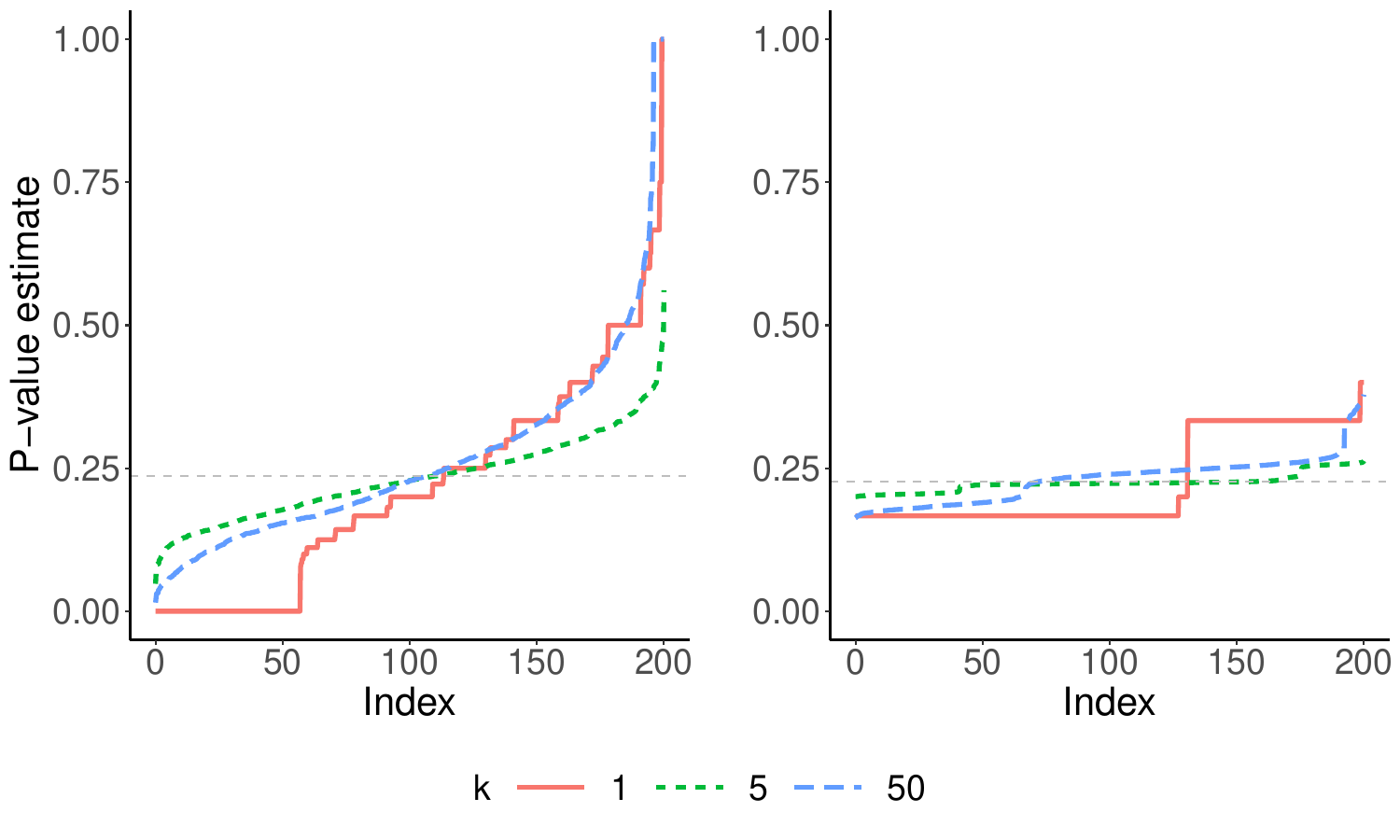}
    \caption{\small{Each plot shows the ordered $p$-value estimates obtained over 1000 runs using importance sampling without stratification (left) and with stratification (right), for a simulated data set with $T = 200$, and using $N = 200$ samples each time. The importance sampling distribution is $Beta(\frac{h}{2k}, \frac{h}{2k})$ where $k = 1, 5, 50$. The dashed grey line corresponds to the mean estimate. The same data set was used for all $p$-value estimates.}}
    \label{fig:strat}
\end{figure}

In Section \ref{sec:lr-psi}, we discussed using stratified sampling to estimate the $p$-value.
Figure \ref{fig:strat} shows plots of the $p$-values we obtained for a given simulated data set, using importance sampling with and without stratification, with 200 samples in each case. (Values of $\phi$ are sampled from $Beta(\frac{h}{2k}, \frac{h}{2k})$ for $k \in \{1, 5, 50\}$; the true distribution of $\phi$ is $Beta(\frac{h}{2}, \frac{h}{2})$.) We can see that although the two methods have the same mean, using stratified sampling results in a much smaller variance in the $p$-value estimate.

\subsection{Extra simulations for Gaussian process approach} \label{sec:finding-l}

For the Gaussian process model, we use a kernel of the form
\begin{equation}
  \label{eq:kernel}
    \boldsymbol{K}(\phi_1, \phi_2) = e^{-\frac{1}{2l^2} |\phi_1 - \phi_2|},
\end{equation}
where $\phi_1, \phi_2 \in [0, 1]$ are two possible values of $\phi$. The value $l$ is a hyperparameter, which must be chosen before we fit the model. A larger $l$ will lead to a smoother function $\hat{p}(\phi)$, as it will more strongly force $\hat{p}(\phi_1)$ and $\hat{p}(\phi_2)$ to take the same value when $\phi_1$ and $\phi_2$ are close together.

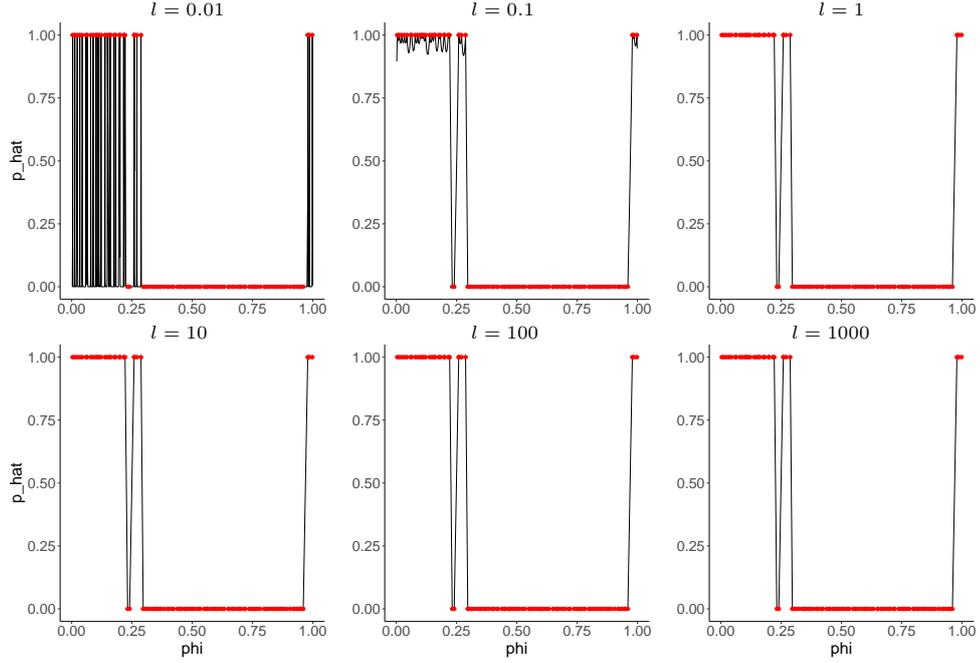
\begin{figure}
    \centering
    \includegraphics[width=0.98\linewidth]{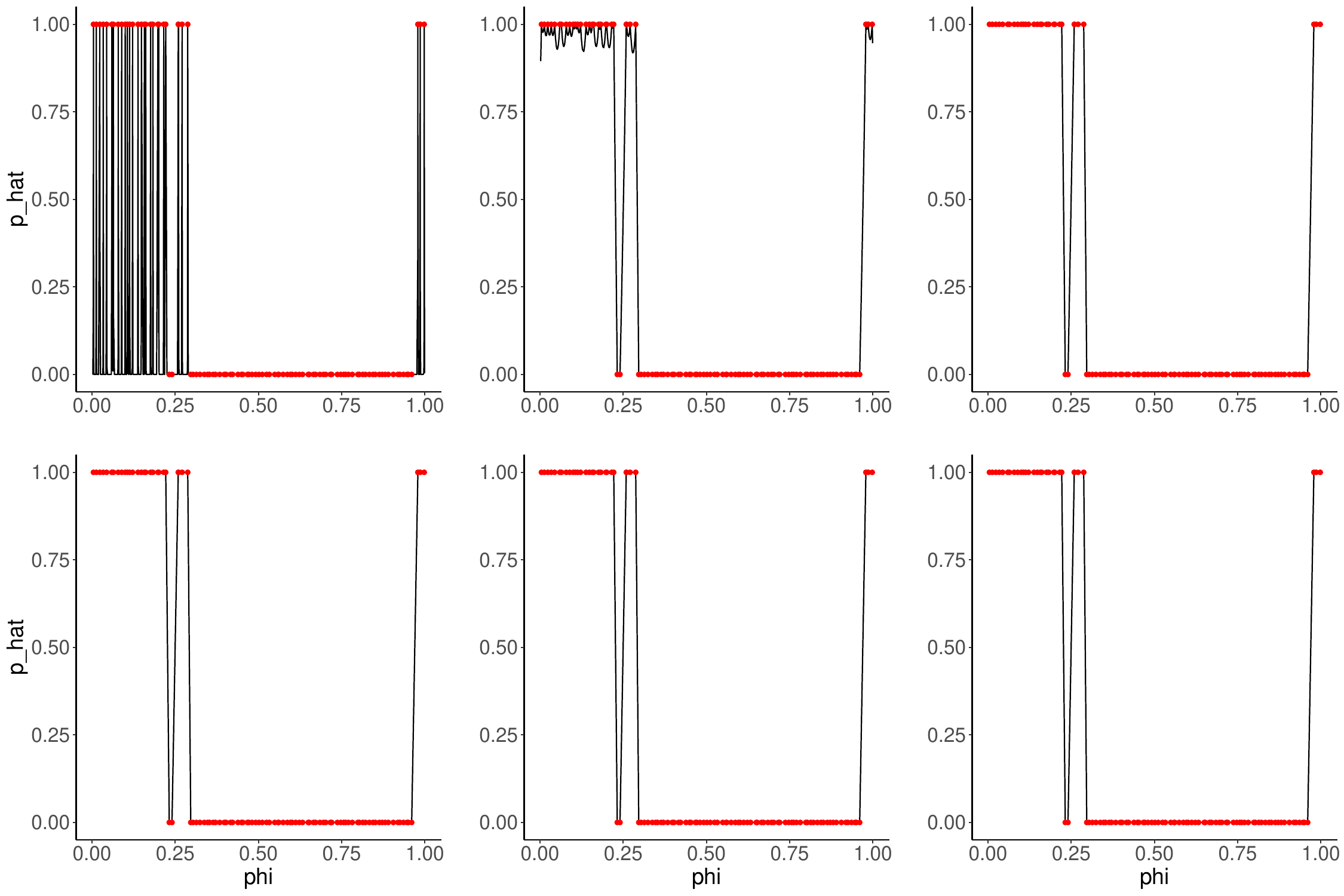}
        \put(-310, 244){\footnotesize{$l = 0.01$}}
        \put(-190, 244){\footnotesize{$l = 0.1$}}
        \put(-60, 244){\footnotesize{$l = 1$}}
        \put(-310, 122){\footnotesize{$l = 10$}}
        \put(-190, 122){\footnotesize{$l = 100$}}
        \put(-70, 122){\footnotesize{$l = 1000$}}
    \caption{\small{Plots showing $\hat{p}(\phi)$ where we fit a Gaussian process with the kernel defined in \eqref{eq:kernel}, where $l = \{0.01, 0.1, 1, 10, 100, 1000\}$.}}
    \label{fig:varying-l}
\end{figure}

In Figure \ref{fig:varying-l}, we investigate the effect of using six different values of $l$ ($l \in \{0.01, 0.1, 1, 10, 100, 1000\}$), for a simulated data set. The figure displays $\hat{p}(\phi)$, shown by a black line, with observed values $p(\phi)$ shown as red points. For $l = 1, 10, 100, 1000$, the estimates we obtain of $p(\phi)$ are very similar, and appropriately smooth: most $\phi$ values have $\hat{p}(\phi)$ in $\{0, 1\}$ (or very close to one of these values), and we expect that $\mathcal{S}$ will consist of a small number of intervals. Using $l = 0.01$ or $l = 0.1$ gives less good results. So, it seems like a value $l \geq 1$ is likely to give good results. In the paper we set $l = 100$ for all simulations and data analyses.

\subsection{PELT}

\begin{figure}
    \includegraphics[width=0.85\linewidth]{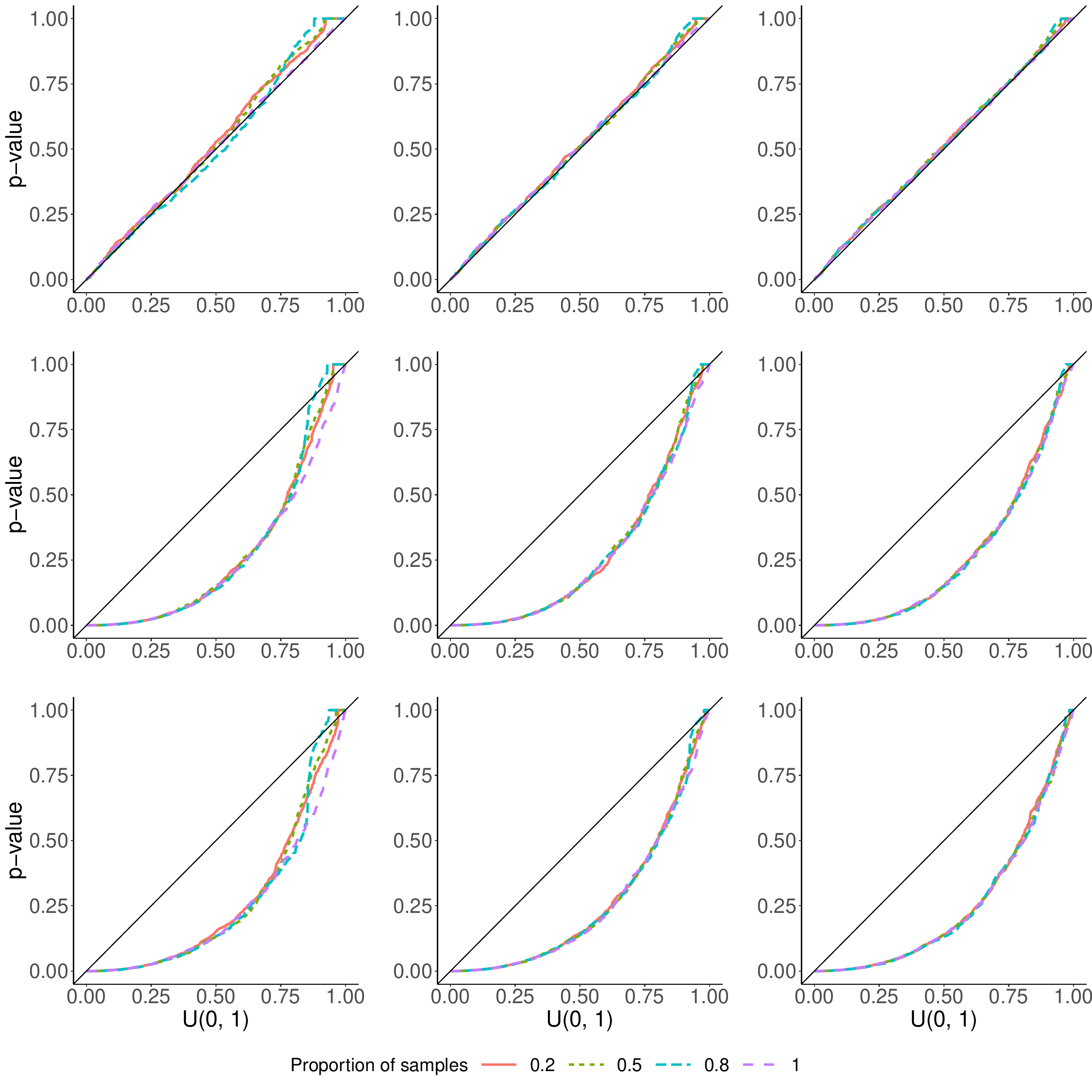}
        \put(-265, 318){\footnotesize{$N = 50$}}
        \put(-158, 318){\footnotesize{$N = 100$}}
        \put(-55, 318){\footnotesize{$N = 200$}}
        \put(0, 270){\footnotesize{$H_0$}}
        \put(0, 180){\footnotesize{$\tau_1 = \frac{T}{4}, \tau_2 = \frac{T}{2}$,}}
        \put(0, 170){\footnotesize{$\tau_3 = \frac{3T}{4}$}}
        \put(0, 80){\footnotesize{$\tau_1, \tau_2, \tau_3$}}
        \put(0, 70){\footnotesize{random}}
    \caption{\small{QQ plots of $p$-values obtained using the likelihood ratio for three scenarios -- no change, three equally spaced changes, and three random changes -- using PELT to estimate changepoints. Each line corresponds to the proportion of samples that were used to fit the Gaussian process: where this is less than 1, the remaining samples were drawn from the posterior of the Gaussian process.}}
    \label{fig:pelt-1}
\end{figure}

Figure \ref{fig:pelt-1} shows QQ plots of $p$-values when simulating under three situations, as in Figure \ref{fig:is-vs-gp} in the main text, but when changepoints are estimated using PELT \citep{killick2012optimal}.

\end{document}